\documentclass[fleqn,usenatbib]{mnras}

\usepackage{newtxtext,newtxmath}
\usepackage[T1]{fontenc}

\DeclareRobustCommand{\VAN}[3]{#2}
\let\VANthebibliography\thebibliography
\def\thebibliography{\DeclareRobustCommand{\VAN}[3]{##3}\VANthebibliography}

\usepackage{graphicx}	% Including figure files
\usepackage{amsmath}	% Advanced maths commands
\usepackage{rotating}
\usepackage{wasysym}    % Extra astronomy symbols

\usepackage{orcidlink}
\usepackage{xspace}
\usepackage{ulem}
\makeatletter % Workaround to only color the year for cite commands
  \patchcmd{\NAT@citex}
    {\@citea\NAT@hyper@{%
      \NAT@nmfmt{\NAT@nm}%
      \hyper@natlinkbreak{\NAT@aysep\NAT@spacechar}{\@citeb\@extra@b@citeb}%
      \NAT@date}}
    {\@citea\NAT@nmfmt{\NAT@nm}%
    \NAT@aysep\NAT@spacechar\NAT@hyper@{\NAT@date}}{}{}

  \patchcmd{\NAT@citex}
    {\@citea\NAT@hyper@{%
      \NAT@nmfmt{\NAT@nm}%
      \hyper@natlinkbreak{\NAT@spacechar\NAT@@open\if*#1*\else#1\NAT@spacechar\fi}%
        {\@citeb\@extra@b@citeb}%
      \NAT@date}}
    {\@citea\NAT@nmfmt{\NAT@nm}%
    \NAT@spacechar\NAT@@open\if*#1*\else#1\NAT@spacechar\fi\NAT@hyper@{\NAT@date}}
    {}{}
\makeatother

\newcommand{\HM}{\ion{H}{$_2$}\xspace}
\newcommand{\HI}{\ion{H}{I}\xspace}
\newcommand{\HII}{\ion{H}{II}\xspace}
\newcommand{\HeI}{\ion{He}{I}\xspace}
\newcommand{\HeII}{\ion{He}{II}\xspace}
\newcommand{\HeIII}{\ion{He}{III}\xspace}

\newcommand{\thesan}{\textsc{thesan}\xspace}
\newcommand{\thzoom}{\mbox{\textsc{thesan-zoom}}\xspace}

\title[The high-redshift star-forming main sequence]{The \thzoom project: Burst, quench, repeat --- unveiling the evolution of high-redshift galaxies along the star-forming main sequence}

\author[W. McClymont et al.]{%
William McClymont $\orcidlink{0009-0009-5565-3790}$,$^{1,2}$\thanks{E-mail: \href{mailto:wjm50@cam.ac.uk}{wjm50@cam.ac.uk} (WM)}
Sandro Tacchella $\orcidlink{0000-0002-8224-4505}$,$^{1,2}$
Aaron Smith $\orcidlink{0000-0002-2838-9033}$,$^{3}$
Rahul Kannan $\orcidlink{0000-0001-6092-2187}$,$^{4}$
\newauthor
Ewald Puchwein $\orcidlink{0000-0001-8778-7587}$,$^{5}$
Josh Borrow  $\orcidlink{0000-0002-1327-1921}$,$^{6}$
Enrico Garaldi $\orcidlink{0000-0002-6021-7020}$,$^{7,8}$
Laura Keating $\orcidlink{0000-0001-5211-1958}$,$^{9}$
Mark Vogelsberger $\orcidlink{0000-0001-8593-7692}$,$^{10}$
\newauthor
Oliver Zier $\orcidlink{0000-0003-1811-8915}$,$^{11,10}$
Xuejian Shen $\orcidlink{0000-0002-6196-823X}$,$^{10}$
Filip Popovic $\orcidlink{0009-0006-8856-918X}$,$^{4}$
and Charlotte Simmonds $\orcidlink{0000-0003-4770-7516}$$^{1,2}$
\\
\\
$^{1}$Kavli Institute for Cosmology, University of Cambridge, Madingley Road, Cambridge CB3 0HA, UK\\
$^{2}$Cavendish Laboratory, University of Cambridge, 19 JJ Thomson Avenue, Cambridge CB3 0HE, UK\\
$^3$ Department of Physics, The University of Texas at Dallas, Richardson, TX 75080, USA \\
$^4$ Department of Physics and Astronomy, York University, 4700 Keele Street, Toronto, ON M3J 1P3, Canada \\
$^5$ Leibniz-Institut f\"ur Astrophysik Potsdam, An der Sternwarte 16, 14482 Potsdam, Germany \\
$^6$ Department of Physics and Astronomy, University of Pennsylvania, 209 South 33rd Street, Philadelphia, PA 19104, USA \\
$^7$ Kavli IPMU (WPI), UTIAS, The University of Tokyo, Kashiwa, Chiba 277-8583, Japan \\
$^8$ Institute for Fundamental Physics of the Universe, via Beirut 2, 34151 Trieste, Italy \\
$^9$ Institute for Astronomy, University of Edinburgh, Blackford Hill, Edinburgh, EH9 3HJ, UK \\
$^{10}$ Department of Physics, Kavli Institute for Astrophysics and Space Research, Massachusetts Institute of Technology, Cambridge, MA 02139, USA \\
$^{11}$ Center for Astrophysics $|$ Harvard $\&$ Smithsonian, 60 Garden Street, Cambridge, MA 02138, USA
}

\date{Accepted XXX. Received YYY; in original form ZZZ}

\pubyear{\the\year{}}

\begin{document}
\label{firstpage}
\pagerange{\pageref{firstpage}--\pageref{lastpage}}
\maketitle

% Abstract of the paper
\begin{abstract}
Characterizing the evolution of the star-forming main sequence (SFMS) at high redshift is crucial to contextualize the observed extreme properties of galaxies in the early Universe. We present an analysis of the SFMS and its scatter in the \thzoom simulations, where we find a redshift evolution of the SFMS normalization scaling as $\propto (1+z)^{2.64\pm0.03}$, significantly stronger than is typically inferred from observations. We can reproduce the flatter observed evolution by filtering out weakly star-forming galaxies, implying that current observational fits are biased due to a missing population of lulling galaxies or overestimated star-formation rates. We also explore star-formation variability using the scatter of galaxies around the SFMS ($\sigma_{\mathrm{MS}}$). At the population level, the scatter around the SFMS increases with cosmic time, driven by the increased importance of long-term environmental effects in regulating star formation at later times. To study short-term star-formation variability, or ``burstiness'', we isolate the scatter on timescales shorter than 50\,Myr. The short-term scatter is larger at higher redshift, indicating that star formation is indeed more bursty in the early Universe. We identify two starburst modes: (i) externally driven, where rapid large-scale inflows trigger and fuel prolonged, extreme star formation episodes, and (ii) internally driven, where cyclical ejection and re-accretion of the interstellar medium in low-mass galaxies drive bursts, even under relatively steady large-scale inflow. Both modes occur at all redshifts, but the increased burstiness of galaxies at higher redshift is due to the increasing prevalence of the more extreme external mode of star formation.
\end{abstract}

% Select between one and six entries from the list of approved keywords.
% Don't make up new ones.
\begin{keywords}
galaxies: high-redshift -- galaxies: ISM -- ISM: lines and bands -- ISM: structure -- cosmology: reionization -- radiative transfer
\end{keywords}

%%%%%%%%%%%%%%%%%%%%%%%%%%%%%%%%%%%%%%%%%%%%%%%%%%

%%%%%%%%%%%%%%%%% BODY OF PAPER %%%%%%%%%%%%%%%%%%

\section{Introduction}
\label{sec:Introduction}

Star formation is a fundamental process in galaxy evolution, through which gas is transformed into stars that shape the physical properties of galaxies over cosmic time. It is a vital ingredient in the chemical enrichment of the interstellar medium \citep[ISM;][]{Lilly:2013aa,Maiolino:2019aa}, the dynamics of galaxies \citep{Wisnioski:2015aa,Hung:2019aa}, and the reionization of the Universe \citep{Robertson:2015aa,Simmonds:2023aa,Simmonds:2024ab,Simmonds:2024aa}. Understanding the mechanisms that regulate star formation is crucial for constructing a comprehensive picture of galaxy formation and evolution \citep{Somerville:2015aa,Naab:2017aa,Vogelsberger:2020aa}.

A key relation in this context is the star-forming main sequence (SFMS), which describes a tight correlation between the star formation rate (SFR) and stellar mass ($\mathrm{M_\ast}$) of star-forming galaxies \citep{Brinchmann:2004aa,Daddi:2007aa,Speagle:2014aa}. The SFMS provides a baseline for understanding the typical growth of galaxies and serves as a reference for identifying peculiar or extreme objects. The shape of the SFMS is well fit by a power law $\propto\mathrm{M}_\ast^\beta$, with most works finding a relatively weak dependence of specific SFR (sSFR$\,\equiv\,$SFR/$M_\ast$) on stellar mass. However, there is disagreement at the high-mass end of the SFMS ($M_\ast\gtrsim10^{10.5}~\mathrm{M_\odot}$), where some works find a turnover \citep{Whitaker:2014aa,Leslie:2020aa,Popesso:2023aa} while others do not \citep{Pearson:2018aa}.

The normalization of the SFMS evolves with redshift, and this dependence can also be described with a power law $\propto(1+z)^\mu$. From simple theoretical arguments based on an analogy to halo growth in dark matter-only (DMO) simulations, the normalization is expected to evolve rapidly, with $\mu\approx5/2$ \citep{Bouche:2010aa,Dekel:2013aa,Tacchella:2016aa}. 
Discrepancies from this expectation would reveal the fingerprints of baryonic physics, from which we could learn about how star formation progresses across cosmic time and depends on factors such as halo mass. Unfortunately, the literature has not reached a consensus, with a wide range of values for $\mu$ quoted and other functional forms proposed \citep{Speagle:2014aa,Whitaker:2014aa,Ilbert:2015aa,Thorne:2021aa,Leja:2022aa}, although some works have undertaken a compilation of literature in an attempt to converge on an understanding of the redshift evolution \citep{Speagle:2014aa,Popesso:2023aa}.

While the normalization of the SFMS reveals the overall progression of cosmic star formation, the scatter around the SFMS provides insights into the variability of star formation, and therefore the underlying physics of galaxy formation and evolution \citep{Iyer:2020aa,Tacchella:2020aa,Fortuntextbackslashe:2025aa}. Particularly helpful in this regard is studying the scatter around the SFMS using SFRs averaged over different timescales \citep{Caplar:2019aa}. Short-term fluctuations ($\lesssim$100\,Myr) in SFR can reveal the immediate effects of feedback processes, such as supernovae and stellar winds, on the ISM \citep{Shin:2023aa}. Over longer timescales ($\gtrsim$100\,Myr), variations can indicate changes in gas accretion rates, mergers, or secular evolutionary processes \citep{White:1978aa,Dekel:2014aa,Somerville:2015aa,Tacchella:2018aa,Behroozi:2019aa,Iyer:2020aa}. Studying this variability helps disentangle the interplay between star formation, feedback, and galactic environments. 

Theoretical works studying the variability of star formation generally find that star formation is bursty in low-mass galaxies due to episodic feedback which expels gas from the ISM \citep{Hopkins:2014aa,Hopkins:2023aa,Hayward:2017aa}, and that star formation is bursty across all masses at high redshift due to highly variable gas accretion and short equilibrium timescales \citep{Angles-Alcazar:2017aa,Faucher-Giguere:2018aa,Tacchella:2020aa}. \citet{Tacchella:2020aa} directly study the impact of star-formation variability on SFMS scatter across a range of masses and redshifts, showing that low-mass and high-redshift galaxies are expected to have a larger scatter around the SFMS.

Observational measures of the SFMS scatter vary wildly in the literature, with no clear consensus on the magnitude of the scatter or the trend with mass and redshift \citep{Guo:2013aa,Pearson:2018aa,Leja:2022aa,Clarke:2024aa,Cole:2025aa}. Various complications arise in measuring this scatter, and a particularly fundamental issue is that measurements of SFMS scatter are plagued by sample selection issues. SFRs are not log-normal at a fixed mass and instead show a long tail down to zero. Due to limitations of observational SFR indicators, this tail can artificially bunch up, appearing as a second quiescent sequence. Often, attempts are made to enforce log-normality by defining ``star-forming galaxies'' to avoid galaxies found in the low SFR tail, however, this of course will bias any measurement of scatter based on the selection criteria used. \citet{Feldmann:2017aa} suggest that the SFMS is fit using zero-inflated negative binomial distributions (zNBDs), which are distributions which can account for a quenching tail and for galaxies with zero ongoing star formation.

The comparison of theoretical and observational work on the SFMS is further complicated by several effects. For example, SFRs from simulations can be directly measured, whereas in observations they must be inferred indirectly from tracers. In particular, star formation is commonly traced on a $\sim$100\,Myr timescale with UV emission and on a $\sim$10\,Myr timescale with H$\alpha$. These tracers have a strong foundation; in star-forming galaxies, the UV emission is dominated by bright, young stars \citep[$\lesssim$100\,Myr;][]{Salim:2007aa}, and the H$\alpha$ emission generally arises from gas which has been ionized by the copious amount of Lyman Continuum (LyC) photons emitted by very young stars \citep[$\lesssim$10\,Myr;][]{Kennicutt:1998aa}. These tracers are affected by biases including uncertainties in the age-metallicity relationship, the nature of the initial mass function (IMF), and assumptions about star formation histories \citep[SFHs;][]{Conroy:2013aa}. The effect of dust attenuation can be dramatic for both H$\alpha$ and UV due to obscured star formation, although this effect mostly impacts massive galaxies \citep[e.g.,][]{Maheson:2024aa,Sandles:2024aa}. While it is in principle possible to correct for this attenuation, there are even further difficulties at high redshift due to the uncertain attenuation curves \citep{Reddy:2025aa}. Additionally, due to complex physical processes such as LyC escape, H$\alpha$ is an imperfect tracer of LyC emission which induces further uncertainties in its use as a SFR indicator \citep{Tacchella:2022aa}. In this work, as well as investigating the SFMS through the lens of the intrinsic star-formation rates, we will use forward-modeled observables to understand how the observed SFMS may differ from the intrinsic SFMS.

The advent of the \textit{James Webb Space Telescope (JWST)} has drastically advanced the redshift frontier for studies of the SFMS \citep{Stark:2025aa}. While there have recently been a number of studies at $z>3$ using data from \textit{JWST} studying the SFMS and its scatter \citep{Clarke:2024aa,Cole:2025aa}, the sample sizes are thus far too small to draw firm conclusions. However, there have been other indications of extreme variability, or burstiness, in high-redshift galaxies \citep{Endsley:2024ab,Endsley:2025aa,Ciesla:2024aa}. These include mini-quenched galaxies \citep{Looser:2024aa,Looser:2025aa,Baker:2025aa,Trussler:2025aa}, galaxies in the process of rapidly quenching \citep{McClymont:2025ad}, and rejuvenating galaxies \citep{Witten:2025aa}. 
Studies of the high-redshift UV luminosity function with \textit{JWST} have revealed an apparent over-abundance of UV-bright galaxies \citep{Donnan:2023aa,Donnan:2024aa,Leethochawalit:2023aa,Perez-Gonzalez:2023aa,Robertson:2024aa,Whitler:2025aa}. Bursty star-formation has been invoked to explain these results by inducing large UV variability which can allow lower mass galaxies to temporarily scatter to high UV luminosity, although the expected UV scatter from bursty star formation is currently unclear \citep{Shen:2023aa,Sun:2023aa,Mason:2023aa,Kravtsov:2024aa}. There have been many recent theoretical works studying bursty star formation, however, a clear consensus on its cause and magnitude has not yet been reached \citep{Faisst:2024aa,Dome:2024aa,Dome:2025aa,Kimmig:2025aa,Gelli:2025aa}. These results indicate that there is much to learn about the nature of star formation in the early Universe and that the SFMS is a promising avenue to explore. 

It is in this context that we analyze the SFMS in the \thzoom simulations \citep{Kannan:2025aa}, which are zoom-in radiation hydrodynamics simulations capturing the muti-phase ISM with high spatial and mass resolution at $z \gtrsim 3$. Our goal for this work is to provide predictions and explanations for the SFMS scatter and normalization at high redshift, and in general to better understand the causes and effects of bursty star formation in the early Universe. We will fit the SFMS across $3<z<12$ and study its slope and normalization in the context of previous theoretical and observational studies, with particular reference to the apparent flattening of the redshift evolution seen in observations. We will aim to understand both the overall variability of star formation on all timescales, as well as more specifically galaxy ``burstiness'', which we define as variability on timescales of $\sim$50\,Myr or less. We will also investigate the causes of burstiness, including internal and external driving mechanisms.

In Section~\ref{sec:Simulation methodology}, we summarize the \thzoom simulations, our galaxy sample, and the post-processing of the simulations to produce synthetic observables. In Section~\ref{sec:Star-forming main sequence}, we fit and analyze the SFMS, including its dependence on stellar mass and redshift. Section~\ref{sec:Main sequence scatter} focuses on the scatter of the main sequence and how it relates to the variability of star formation. In Section~\ref{sec:Conclusions}, we briefly summarize our findings.

\section{Simulation methodology}
\label{sec:Simulation methodology}

\subsection{Simulations}
\label{sec:Simulations}

\begin{figure*} 
\centering
	\includegraphics[width=\textwidth]{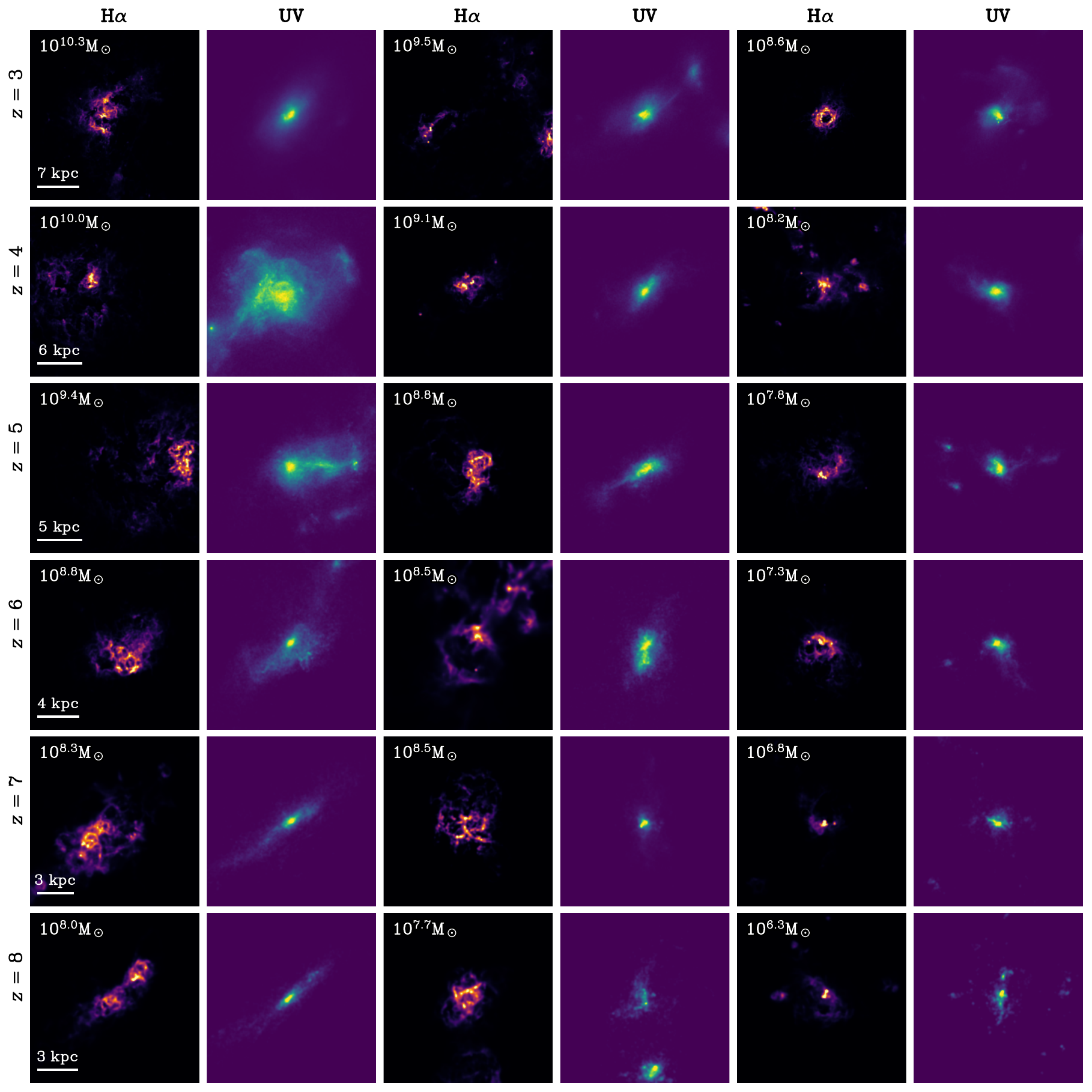}
    \caption{Three massive \thzoom galaxies, m12.2 subhalo 0 (the two left columns), m11.5 subhalo 0 (the two central columns), and m11.1 subhalo 0 (the two right columns) in H$\alpha$ ($1^\text{st}$, $3^\text{rd}$, and $5^\text{th}$ column) and UV ($2^\text{th}$, $4^\text{th}$, and $6^\text{th}$ column) emission from $z=3$ to $z=8$ top to bottom. All images are 120\,ckpc by 120\,ckpc, and a bar in the lower left of each row shows the scale in pkpc. Luminosity is scaled between zero and the 99.9th percentile brightest pixel in an image. The galaxies show remarkable variation across cosmic time, including distortions due to galaxy--galaxy interactions. Offsets between the UV and H$\alpha$ emission can arise due to a variety of effects, including radiative transfer processes and sites of star formation being spatially disconnected from older stars \citep{McClymont:2025ab}.}
    \label{fig:simple_images}
\end{figure*}

This work presents an analysis of the \thzoom simulation suite \citep{Kannan:2025aa}. These simulations are a high-resolution extension of the \thesan project \citep{Kannan:2022aa, Smith:2022ab, Garaldi:2022aa}, which were large-volume ($\sim$100\,cMpc) simulations using the IllustrisTNG galaxy formation model \citep{Pillepich:2018aa} and on-the-fly radiative transfer. The original \thesan simulations aimed to study galaxy formation and self-consistently model the Epoch of Reionisation (EoR), with ionising photons being sourced from stars and black holes. While this ambitious project has seen extensive use \citep[e.g.][]{Yeh:2023aa,Neyer:2024aa,Garaldi:2024aa,Shen:2024aa,Jamieson:2025aa}, it is not well suited to tackle key astrophysical questions about small-scale star formation and the ISM due to the effective equation-of-state galaxy formation approach \citep[e.g., ][]{Springel:2003aa,Vogelsberger:2013aa,Pillepich:2018aa}.

The \thzoom simulations aim to combine the key advantages of the \thesan approach with a detailed model of galaxy formation, including a multi-phase ISM and local stellar feedback. The zoom regions are extracted from the parent \thesan volume, enabling the study of galaxy formation processes within their larger cosmological context, uniquely including the radiation field from the original simulations as boundary conditions. The realistic, time-varying radiation environment is vital for studying high-redshift galaxies, where radiative feedback from nearby galaxies is expected to be important \citep{Rosdahl:2018aa,Ocvirk:2020aa}. Initial works which have been conducted using \thzoom have studied galaxy-scale star-formation efficiencies \citep[SFEs;][]{Shen:2025aa}, the impact of reionisation on galaxies \citep{Zier:2025aa}, Population III star formation \citep{Zier:2025ab}, high-redshift galaxy sizes \citep{McClymont:2025ab}, SFEs on the scale of giant molecular clouds \citep[GMCs;][]{Wang:2025aa}, and chemical evolution \citep{McClymont:2025ae}.

Full details of the simulation are provided in \citet{Kannan:2025aa}, however we outline some key information here. The \thzoom simulations are performed using the {\sc arepo-rt} code \cite{Kannan:2019aa}, a radiation hydrodynamics extension of the moving-mesh code {\sc arepo} \citep{Springel:2010aa}, including the recently implemented node-to-node communication strategy \citep{Zier:2024aa}. The radiation transfer is treated on-the-fly using a moment-based approach, which tracks the photon number density and flux with a reduced speed of light approximation to increase computational efficiency. This allows for a detailed, spatially resolved modeling of the radiation field, which is crucial for capturing the interaction between ionizing radiation and the ISM. The non-equilibrium thermochemical network tracks six key chemical species—$\HM, \HI, \HII, \HeI, \HeII,$ and $\HeIII$. Metal cooling is implemented assuming ionization equilibrium with a \cite{Faucher-Giguere:2009aa} UV background, and in practice the rates are pre-calculated in \textsc{cloudy} \citep{Ferland:2017aa} and stored in look-up tables \citep{Vogelsberger:2013aa}.

There are three zoom levels, representing factors of 4, 8, and 16 improvements in the spatial resolution (named ``4x'', ``8x'', and ``16x''), which correspond to factors of 64, 512, and 4096 in the mass resolution. This means the zooms have baryonic mass resolutions of $9.09\times10^3\,\mathrm{M}_\odot$, $1.14\times10^3\,\mathrm{M}_\odot$, and $1.42\times10^2\,\mathrm{M}_\odot$ respectively. In this work we always use fiducial physics runs and do not consider model variations. Zoom regions were run with differing resolution levels so we select the highest resolution run available for a given object.

The simulations include a sophisticated stellar feedback model that incorporates photoionization, radiation pressure, stellar winds, and supernova feedback \citep{Kannan:2020aa, Kannan:2021aa}. Stellar radiation is treated self-consistently, with photons emitted by young stars contributing to local heating and ionization. Supernova feedback is implemented by injecting thermal energy and momentum into the surrounding gas. Stellar winds follow the prescription of the SMUGGLE model described in \citet{Marinacci:2019aa}. These feedback mechanisms regulate star formation and drive outflows, which play a crucial role in shaping the properties of early galaxies. Additionally, early stellar feedback is implemented to prevent excessive star formation in dense regions by disrupting molecular clouds shortly after the onset of star formation.

The \thzoom simulations are ideal for studying the star-forming main sequence at high redshift, as they provide detailed information on gas dynamics, star formation, and feedback in the context of a dynamically evolving radiation environment. For each subhalo, we calculate stellar masses and SFRs based on bound particles within the virial radius, which is defined with the spherical overdensity criterion $R_\mathrm{vir}=R_{\rm crit,200}$. SFRs are calculated based on the number of stars formed within the given averaging timescale ($t_\mathrm{avg}$) as
\begin{equation}
  \mathrm{SFR}_{t_\mathrm{avg}} = \frac{\sum_im_{\ast,i}}{t_\mathrm{avg}}\,,
\end{equation}
where $m_{\ast,i}$ is the initial mass of a stellar particle. In this work, we consider only subhalos which are resolved with at least 1000 stellar particles to allow us to resolve a minimum sSFR over 10\,Myr of $\mathrm{sSFR_{10}}=0.1\,\mathrm{Gyr^{-1}}$. This is in effect a stellar mass cut of $9.09\times10^6\,\mathrm{M}_\odot$, $1.14\times10^6\,\mathrm{M}_\odot$, and $1.42\times10^5\,\mathrm{M}_\odot$ for the 4x, 8x, and 16x resolutions respectively, though this does not account for mass loss. To trace the inflow history, we further require that the subhalo can be found as a progenitor in the merger tree of a subhalo which is uncontaminated by low-resolution particles in the final snapshot. Haloes were identified with the friends-of-friends (FOF) algorithm \citep{Davis:1985aa}, and self-gravitating subhalos were identified within the FOF groups using the SUBFIND-HBT algorithm \citep{Springel:2001aa,Springel:2021aa}. We include both central and satellite galaxies (subhaloes) in this analysis. We also only consider galaxies in the redshift range $3<z<12$. This leaves us with 30222 subhalos, comprised of 575 unique trees.

\subsection{Radiative transfer of non-ionising photons}
\label{sec:Radiative transfer of non-ionising continuum and emission line photons}

In this work we are interested in the H$\alpha$ and UV emission of galaxies and the SFRs derived from these observational probes. We compute the continuum and emission luminosities using the Monte Carlo radiative transfer (MCRT) Cosmic Ly$\alpha$ Transfer code \citep[\textsc{colt};][]{Smith:2015aa, Smith:2019aa, Smith:2022aa}. Images of H$\alpha$ and UV emission for \thzoom galaxies are shown in Fig.~\ref{fig:simple_images}. The procedure is largely as described in previous works \citep[e.g.][]{Smith:2022aa,Tacchella:2022aa,McClymont:2024aa}, however we briefly summarize the implementation here.

We first perform MCRT of ionising photons in order to solve for the distribution ionisation states. We choose to use this method rather than the on-the-fly states due to the excessively high H$\alpha$ emission which can arise from numerical issues such as unresolved Str\"{o}mgren spheres \citep[][]{Smith:2022aa} The ionising continuum is sampled from the stellar SEDs, which are derived from the Binary Population and Spectral Synthesis (BPASS) model including binaries \citep[v2.2.1;][]{Eldridge:2009aa, Eldridge:2017aa}\footnote{Further information on BPASS can be found on the project website at \href{https://bpass.auckland.ac.nz}{\texttt{bpass.auckland.ac.nz}}.}. We use a Chabrier IMF \citep{Chabrier:2003aa} with a maximum stellar mass of $100\,\mathrm{M_\odot}$.

H$\alpha$ is a recombination line, and so the luminosity of a given gas cell is given by
\begin{equation}
  L_\mathrm{H\alpha}^\mathrm{rec} = h \nu_\mathrm{H\alpha} \int P_{\mathrm{B},\mathrm{H\alpha}}(T,n_{\rm e}) \alpha_\mathrm{B}(T)\,n_{\rm e} n_\HII\,\text{d}V \, ,
\end{equation}
where the energy at line center is $h\nu_\mathrm{H\alpha}$, and $n_{\rm e}$ and $n_\HII$ are the number densities for free electrons and \HII, respectively. $\alpha_\mathrm{B}$ is the Case B recombination coefficient, and $P_{\mathrm{B},\mathrm{H\alpha}}$ is the probability for that emission line photon to be emitted per recombination event and is taken from \citet{Storey:1995aa}. The coefficient $\alpha_\mathrm{B}$ is from \citet{Hui:1997aa}. A temperature floor of 7000\,K is applied to $P_{\mathrm{B},\mathrm{H\alpha}}$ in order to avoid nonphysical Balmer line ratios due to artificially cold ionized gas \citep[see][]{Smith:2022aa}. This temperature floor is not applied to the recombination rate.

Continuum emission is sourced from the stellar SEDs. We are interested in FUV emission, so we take the average over a wavelength window of 1475--1525\AA. We only include the stellar contribution to the continuum, however we note that the nebular continuum can make a significant contribution to high-redshift galaxies \citep[e.g.][]{Cameron:2024aa,Katz:2025aa}, although this is usually subdominant in the FUV \citep{Tacchella:2025aa}. Photon packets are sampled based on the luminosity and assigned a given weight. Photons are launched isotopically from the source, which is either a star or a random position within the gas cell, for continuum and line emission respectively.

\section{Star-forming main sequence}
\label{sec:Star-forming main sequence}

\subsection{Measuring the star-forming main sequence}
\label{sec:Measuring the star-forming main sequence}

\begin{table}
    \centering
    \begin{tabular}{ccccc}
        \hline
        \multicolumn{5}{|c|}{Redshift and mass-dependent SFMS fit} \\
        \hline
        Tracer & $\mathrm{s_b}$ & $\beta$ & $\mu$ & \\
        SFR$_{10}$ & $0.033\pm0.002$ & $0.041\pm0.004$ & $2.64\pm0.03$ & \\
        SFR$_{30}$ & $0.037\pm0.002$ & $0.042\pm0.004$ & $2.57\pm0.03$ & \\
        SFR$_{50}$ & $0.043\pm0.002$ & $0.041\pm0.004$ & $2.47\pm0.03$ & \\
        SFR$_{100}$ & $0.067\pm0.002$ & $0.032\pm0.002$ & $2.19\pm0.02$ & \\
        SFR$_{\mathrm{H}\alpha}$ & $0.048\pm0.009$ & $0.077\pm0.015$ & $2.60\pm0.07$ & \\
        SFR$_{\mathrm{UV}}$ & $0.051\pm0.001$ & $0.022\pm0.001$ & $2.32\pm0.02$ & \\
        \hline
    \end{tabular}
    \caption{The best fit parameters for Eq.~(\ref{eq:sfms_tacc}). The fitting was carried out with a zNBD likelihood (see \citealt{Feldmann:2017aa}). SFR$_{\mathrm{H}\alpha}$ and SFR$_{\mathrm{UV}}$ were calculated using our calibrations, Eqs.~(\ref{eq:sfr8}) and (\ref{eq:sfr24}), respectively. Errors were obtained through bootstrapping. The normalisation of the SFMS at short timescales increases more rapidly than the normalisation at longer averaging timescales, which can be attributed to galaxies at higher redshift being more likely to have rising SFHs.}
    \label{tab:sfms_fits}
\end{table}

\begin{figure*} 
\centering
	\includegraphics[width=\textwidth]{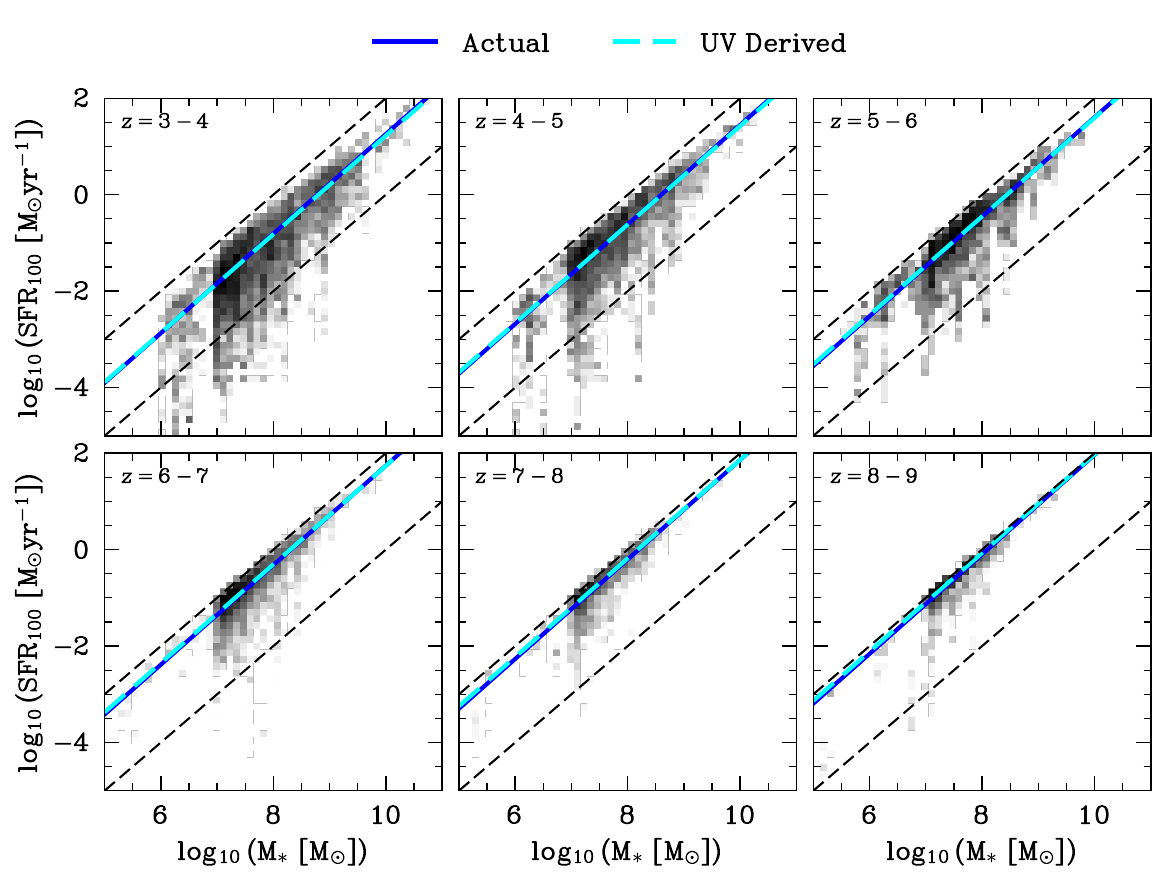}
    \caption{The star-forming main sequence (SFMS) for the \thzoom galaxies across six redshift bins. The shading shows the log-scale number density of galaxies. The solid blue line represents our best fit to the SFMS using the intrinsic values for SFR$_{100}$, whereas the cyan dashed lines show our best fit to the SFR$_\mathrm{UV}$ derived from the intrinsic UV emission (1500\AA). The upper dashed black line shows the one-to-one line, corresponding to all stars having formed within the averaging time, whereas the lower dashed black line corresponds to the value of $m_{\mathrm{sSFC}}$ used in the fitting. Our best fit to the intrinsic SFMS$_{100}$ shows a lower redshift dependence than for SFMS$_{10}$, which is likely due to the bunching up of galaxies on the one-to-one line at high redshift for the 100\,Myr. An alternative way to view this effect is that most galaxies have rising SFHs at high redshift, whereas there is an equilibrium of rising and falling SFHs at lower redshift. This requires that the shorter-timescale SFRs must evolve more rapidly with redshift. The apparent agreement between SFMS$_\mathrm{UV}$ and SFMS$_\mathrm{100}$ is actually serendipitous. The UV best traces SFR$_{24}$ which evolves more rapidly with redshift, but is also biased to a flatter redshift evolution by various effects (see text for details).}
    \label{fig:sfms}
\end{figure*}

\begin{figure*} 
\centering
	\includegraphics[width=\textwidth]{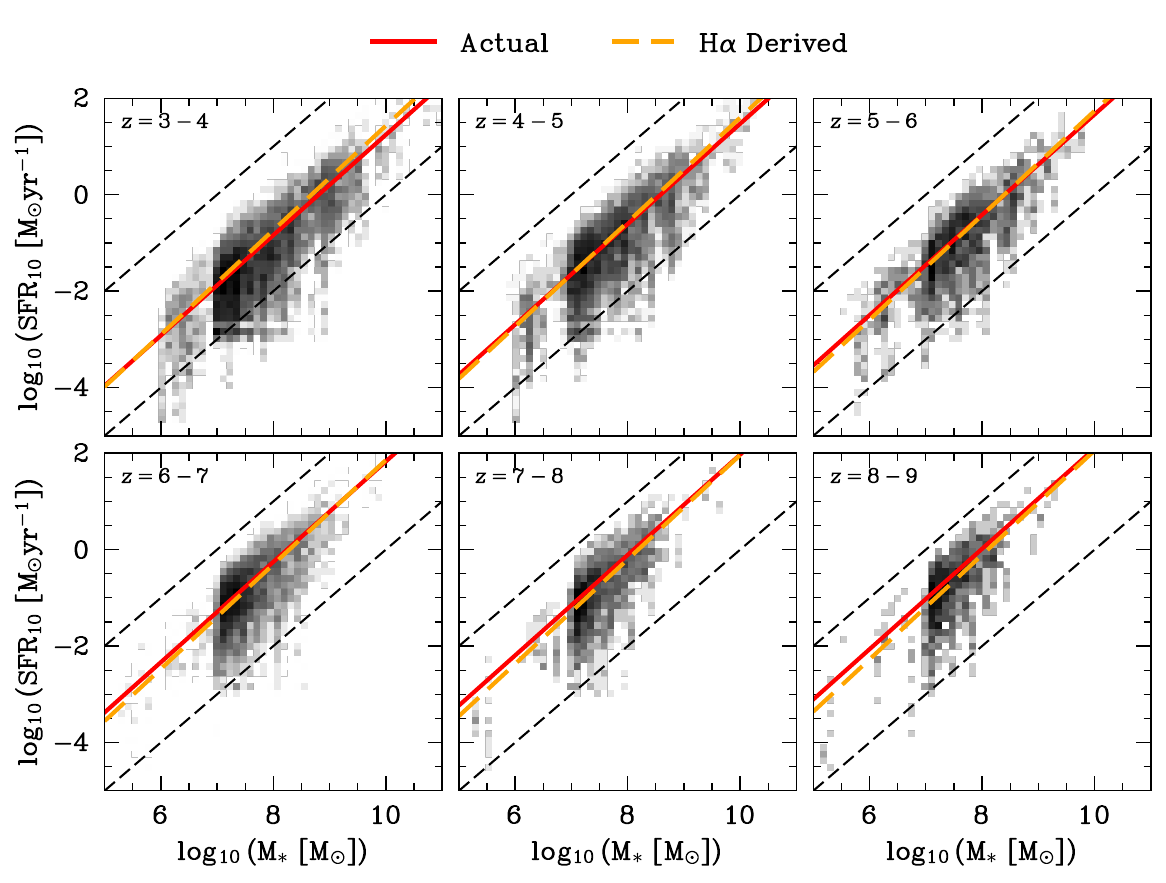}
    \caption{The same as Fig.~\ref{fig:sfms} but for SFR$_{10}$ and SFR$_{\mathrm{H}\alpha}$ derived from the intrinsic H$\alpha$ emission. Our SFMS$_{10}$ fit shows remarkably good agreement with expectations from dark matter halo growth in DMO simulations. This implies that overall galaxy evolution is largely regulated and described by hierarchical structure growth despite the myriad baryonic processes included in our simulations.}
    \label{fig:sfms_ha}
\end{figure*}

The SFMS is often fit with a simple linear regression of stellar mass versus star-formation rate in redshift bins. However, the SFMS is also dependent on redshift, and so fitting in redshift bins includes scatter due to the redshift evolution within each bin \citep{Speagle:2014aa}. We focus our analysis instead on the SFMS parametrization as in \citet{Tacchella:2016aa}. This parametrization is based on the redshift-dependent halo accretion rate, which is approximated as,
\begin{equation}
\frac{\dot{M}_\mathrm{h}}{M_\mathrm{h}}\simeq\mathrm{s_b} \left( \frac{M_\mathrm{h}}{10^{12}~\mathrm{M}_\odot} \right)^\beta (1+z)^\mu~\mathrm{Gyr}^{-1}\,,
\end{equation}
where $s_b$ is the normalization, $\beta$ is the power law mass dependence, and $\mu$ is the power law redshift dependence. At $z>1$, a $\Lambda$CDM universe is in the Einstein-deSitter regime and it can be shown that $\mu\rightarrow5/2$ \citep{Dekel:2013aa}. The halo mass dependence is due to the log-slope of the power spectrum on galactic scales \citep{Dekel:2013aa}. \citet{Neistein:2008aa} showed that $\beta=0.14$ using Millennium simulations halos with masses in the range $10^{11}-10^{14}\mathrm{M}_\odot$. It is important to note, however, that different DMO simulations find somewhat different results for the halo mass and redshift dependence of the accretion rate, especially amongst simulations focused on different halo mass and redshift regimes \citep{Rodriguez-Puebla:2016ab,Yung:2024aa}. \citet{Tacchella:2016aa} use the same functional form to fit the SFMS,
\begin{equation}
\label{eq:sfms_tacc}
\mathrm{sSFR_{MS}}(M_\ast,z)=\mathrm{s_b} \left( \frac{M_\ast}{10^{10}~\mathrm{M}_\odot} \right)^\beta (1+z)^\mu~\mathrm{Gyr}^{-1}\,.
\end{equation}
This parametrization allows us to fit all galaxies simultaneously, and also to interpret our results relative to the expected values based pure on dark matter halo growth. We fit this main sequence using star-formation rates for a variety of timescales and as derived from H$\alpha$ and UV. At a fixed redshift and mass, the SFRs of our galaxies show a long tail toward zero SFR, and so assuming a log-normal distribution can lead to biased results. We instead follow \citet{Feldmann:2017aa}, who showed that the SFMS scatter is well described by zero-inflated negative binomial distributions (zNBDs). zNDBs are able to account for both the quenching tail of the SFMS and for an excess of galaxies with zero ongoing star formation. We therefore fit each timescale by minimizing a negative log-likelihood defined by a zNBD where the expected counts are defined by Eq.~(\ref{eq:sfms_tacc}). We fit excess zero-count probability ($\pi$) as a constant number across mass and redshift. The shape parameter ($\theta$) is related to the SFMS scatter as $\theta=\sigma_\mathrm{MS}/\ln(10)$ where $1\leq\theta\ll\mu_{\mathrm{F17}}$\footnote{We use $\mu_{\mathrm{F17}}$ to distinguish the $\mu$ defined in \citet{Feldmann:2017aa} from the redshift evolution of the SFMS.} \citep{Feldmann:2017aa}. We approximate $\sigma_\mathrm{MS}$ as a broken power law redshift dependence and single power law mass dependence to be fit
\begin{equation}
\label{eq:theta}
\sigma_\mathrm{MS} = \sigma_0 \left( \frac{M_\ast}{10^{10}~\mathrm{M}_\odot} \right)^\gamma \left(\frac{1+z}{z_b}\right)^{-\alpha_1}\left[\frac{1}{2}+\left(\frac{1+z}{2z_b}\right)^{1/\Delta}\right]^{(\alpha_1-\alpha_2)\Delta}\,,
\end{equation}
where $\sigma_0$ is a normalisation, $\gamma$ is the power law slope for mass, $\alpha_1$ and $\alpha_2$ are the power law slopes for redshift, $z_b$ is the break parameter, and $\Delta$ is the smoothness of the break. We note that if we instead approximate $\theta$ as constant, our values of $\mathrm{s_b}$, $\beta$, and $\mu$ are nearly identical, however, this approach allows us to have best-fit functions of  $\sigma_\mathrm{MS}$ which we use in Section~\ref{sec:Main sequence scatter}. The free parameters for each fit are $\mathrm{s_b}$, $\beta$, $\mu$, $\theta$, $\sigma_0$, $\gamma$, $\alpha_1$, $\alpha_2$, $z_b$, $\Delta$, and $\pi$.

For each fit, we perform the minimization 50 times with random parameter initializations and take the best fit to avoid local minima. Uncertainties are estimated via the bootstrap method, where we resample our galaxies with replacement 1000 times and perform 50 minimizations for each with random parameter initializations (50000 total minimizations to estimate uncertainties for each fit). zNBDs are discrete distributions, which means we must convert our sSFRs to integer counts of a specific-mass, $m_{\mathrm{sSFC}}$, which represents a minimum star-formation event. We select $m_{\mathrm{sSFC}}$ to be the minimum sSFR which we can resolve in the smallest subhalo for the shortest timescale we consider for our SFMS fits, 10\,Myr. The smallest subhaloes we consider are resolved with 1000 stellar particles, and the minimum amount of star-formation we could resolve is 1 stellar particle, we use $m_{\mathrm{sSFC}}=1000/t_\mathrm{avg,\,min}\,\,\mathrm{Gyr^{-1}}=0.1\,\,\mathrm{Gyr^{-1}}$. While we consider this the most well-motivated choice for $m_{\mathrm{sSFC}}$, it does not have a strong impact on the fits and our results are encouragingly robust to even unreasonable $m_{\mathrm{sSFC}}$ values\footnote{Even for an order of magnitude increase or decrease in $m_{\mathrm{sSFC}}$, we recover fit parameters which are within the quoted bootstrap errors in Tab.~\ref{tab:sfms_fits}.}. The results from a subset of our fits are shown in Tab.~\ref{tab:sfms_fits}. Fig.~\ref{fig:sfms} shows our SFMS fit for SFR$_{100}$ and SFR$_{\mathrm{UV}}$ compared to our data and fits from literature. Fig.~\ref{fig:sfms_ha} shows the same for SFR$_{10}$ and SFR$_{\mathrm{H\alpha}}$. For completeness, we discuss fits to the SFMS for only central galaxies in Appendix~\ref{sec:Impact of centrals and satellites}.

In our fits to the SFMS, we find that overall the redshift and stellar mass dependence is close to the DMO expectation, with $\mu=2.64\pm0.03$ and $\beta=0.041\pm0.004$ for SFMS$_{10}$ compared to the expected $\mu=2.5$ and $\beta=0.14$. This relatively good agreement of our SFMS fits, which are derived from simulations with state-of-the-art baryonic physics, to expectations derived from DMO simulations is remarkable and demonstrates that much of the physics of galaxy evolution is well described by hierarchical structure growth.

However, our results are not in perfect agreement which may indicate that they do bear the fingerprints of baryonic physics. The redshift dependence is stronger in our fits and the stellar mass dependence is weaker compared to the expectation from dark matter halo growth from \citet{Dekel:2013aa} and \citet{Neistein:2008aa}. These discrepancies may hint at suppressed star formation at later times and in higher-mass haloes. However, we cannot draw strong conclusions given that other DMO simulations find a more rapid redshift evolution and flatter mass evolution for high-redshift and low-mass halos \citep{Rodriguez-Puebla:2016ab,Yung:2024aa}. Interestingly, observational studies of the SFMS have found a flat slope at lower masses which then turns over at higher masses \citep{Popesso:2023aa}. The turnover mass at $M_\ast\approx10^{10.5}\mathrm{M}_\odot$ is above the vast majority of our sample, however it may be related to the suppressed star-formation rates we already find in our higher mass galaxies. 

\subsection{Observational tracers of star formation}
\label{sec:Observational dependence}

\begin{figure} 
\centering
	\includegraphics[width=0.9\columnwidth]{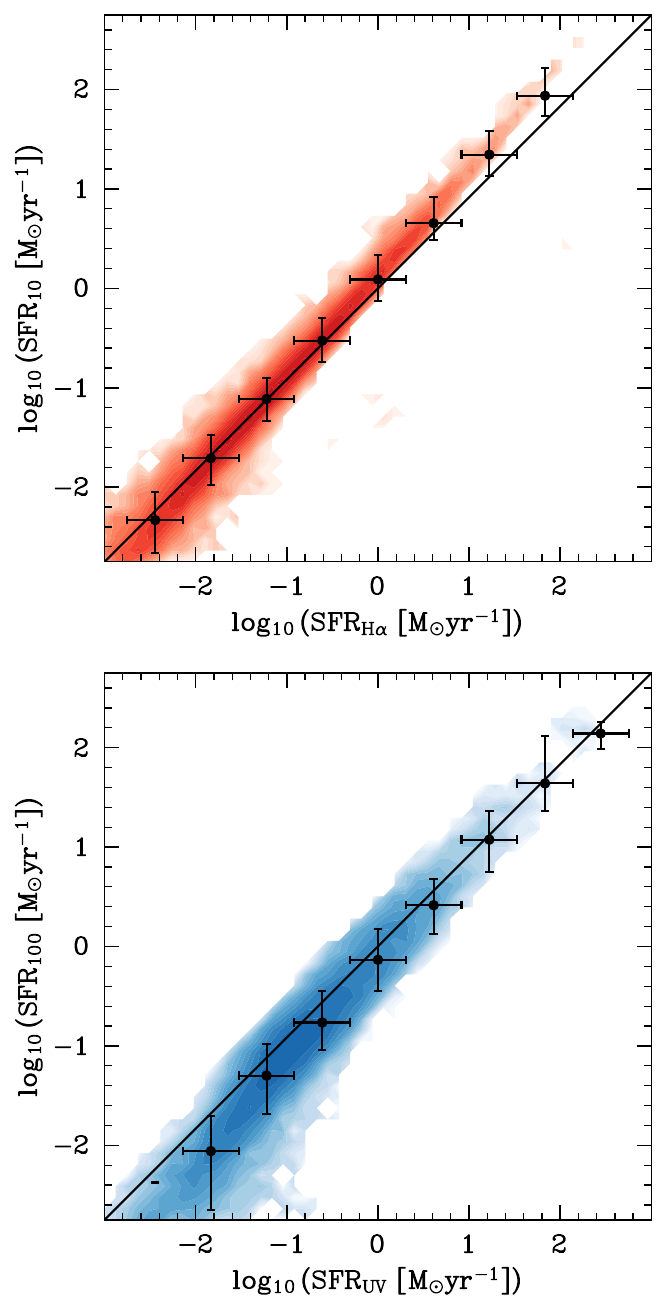}
    \caption{The relationship between intrinsic and observed SFRs. The observed SFRs use traditional SFR calibrations for H$\alpha$ (top panel) and UV (1500\AA, bottom panel). The black line shows the one-to-one line, where all galaxies would lie if the calibration was perfect. The black points show the median values calculated in bins, with the y-axis errors showing the $16^\text{th}$--$84^\text{th}$ range, and the x-axis errors showing the bin range. SFR$_{\mathrm{H}\alpha}$ is systematically lower than SFR$_{10}$ because effects such as LyC escape and dust absorption of LyC tend to decrease H$\alpha$ flux as well as causing significant scatter. SFR$_{\mathrm{UV}}$ is systematically higher than SFR$_{100}$ and is highly scattered, primarily due to the star-formation histories varying on shorter timescales than 100\,Myr.}
    \label{fig:sfrint_sfrobs}
\end{figure}

\begin{figure} 
\centering
	\includegraphics[width=\columnwidth]{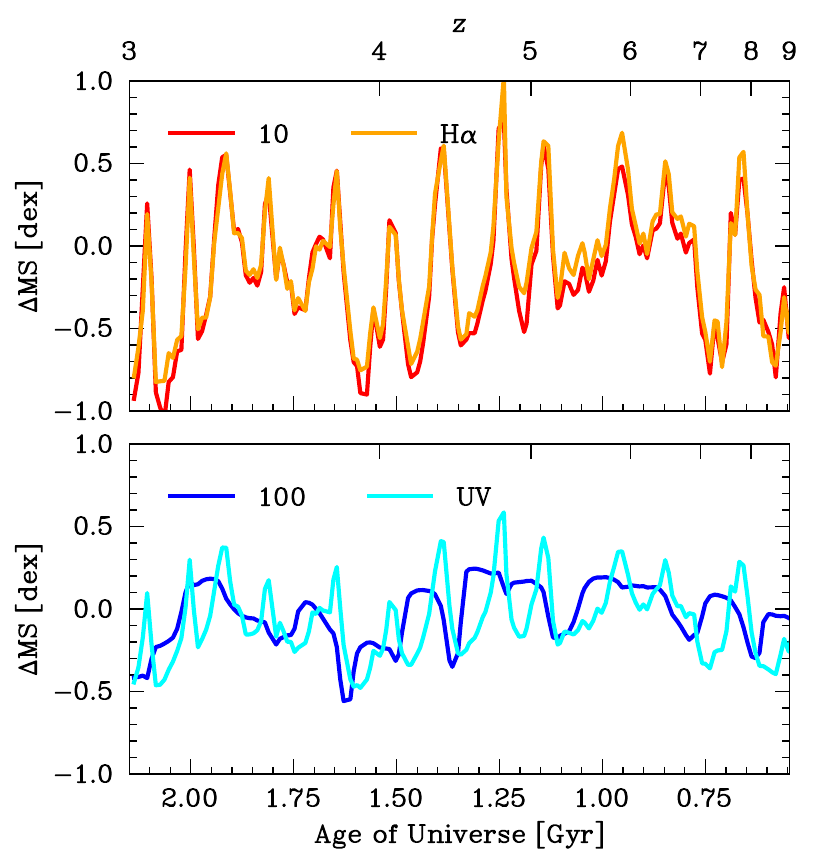}
    \caption{The evolution of SFR offsets from the main sequence ($\Delta\mathrm{MS}$) for an individual galaxy (m12.2, subhalo 0). The offsets from the intrinsic main sequences behave as expected: $\Delta\mathrm{MS}_{100}$ acts as a damped and lagging companion to $\Delta\mathrm{MS}_{10}$. Galaxies with high $\Delta\mathrm{MS}_{10}$ and low $\Delta\mathrm{MS}_{100}$ are beginning a burst, and those with high $\Delta\mathrm{MS}_{100}$ and low $\Delta\mathrm{MS}_{10}$ are recently quenched. The observed picture is more complex. $\Delta\mathrm{MS}_{\mathrm{H}\alpha}$ traces $\Delta\mathrm{MS}_{10}$ well because systemic issues in the SFR$_{\mathrm{H}\alpha}$ to SFR$_{10}$ conversion are less important for SFMS offsets compared to absolute SFR values. However, $\Delta\mathrm{MS}_{\mathrm{UV}}$ (1500\AA) traces $\Delta\mathrm{MS}_{100}$ poorly. This is primarily due to the short nature of the bursts, which leads to a rapidly changing UV to SFR$_{100}$ conversion factor as generations of stars age.}
    \label{fig:evo}
\end{figure}

\begin{figure} 
\centering
	\includegraphics[width=\columnwidth]{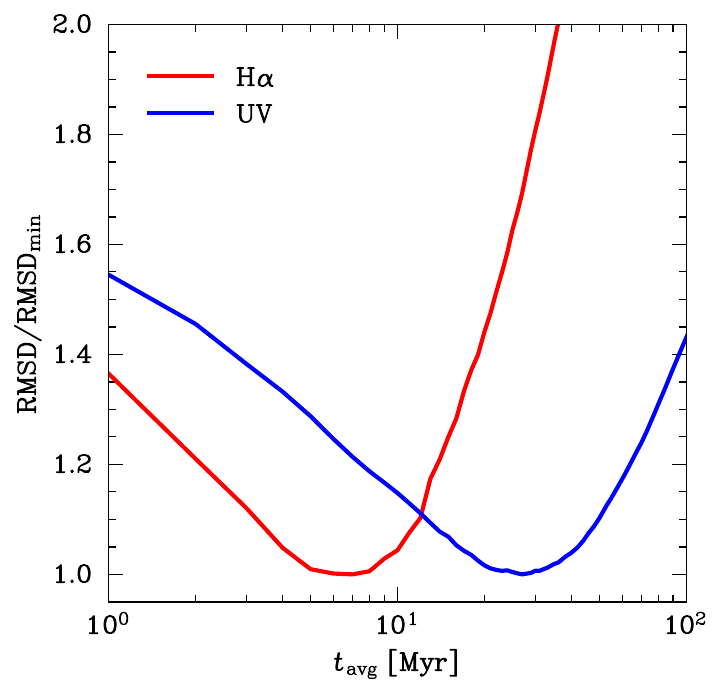}
    \caption{The root mean square deviation (RMSD, Eq.~\ref{eq:rmsd}) as a function of star-formation averaging timescale ($t_\mathrm{avg}$), normalized by the minimum value. The minimum value represents the timescale on which the tracer most effectively probes star formation. H$\alpha$ traces star formation most effectively at 5-8\,Myr, whereas UV (1500\AA)  effectively traces timescales of 22--31\,Myr. The UV traces such short timescales due to the rapidly varying SFRs of high-redshift and low-mass galaxies meaning that the brightest, youngest stars tend to dominate the UV emission. The H$\alpha$ timescale is similarly dependent on the SFH, however other effects, such as the dust obscuration of young stars can be dominant.}
    \label{fig:rmsd}
\end{figure}

In order to investigate observational biases in the measurement of the SFMS, we calculate the intrinsic H$\alpha$ and UV luminosities of each galaxy with COLT (see Section~\ref{sec:Radiative transfer of non-ionising continuum and emission line photons}) and convert them into ``observationally derived'' star-formation rates. The conversion between UV luminosity and star-formation rate is dependent on the stellar metallicity, age, and IMF. We will first make use of commonly used conversions before later deriving our own. Initially, we use the \citet{Clarke:2024aa} rates converted from \citet{Hao:2011aa}, which represents the conversion for a steadily forming generation of stars over 100\,Myr following a \citet{Chabrier:2003aa} IMF with upper mass cutoff of 100\,M$_\odot$\footnote{Specifically, for our tests of commonly used conversions, we use conversion factors of $C=-43.46$\,dex and $C=-41.59$\,dex for UV and H$\alpha$ luminosity, respectively \citep[see][]{Clarke:2024aa}.}.%,

%\begin{equation}
%\mathrm{SFR_{UV}}=\left( \frac{\mathrm{\nu L_{\nu,1500}}}{\mathrm{erg\,s^{-1}}}\right) \times 10^{-42.58} \,\mathrm{M_\odot\,yr^{-1}}.
%\end{equation}

The link between H$\alpha$ luminosity and star-formation rate is based on the use of H$\alpha$ as a number counter for LyC photons, and so the conversion rate is also dependent upon stellar metallicity, age, and IMF. The conversion rate is also dependent upon other factors, such as the LyC escape fraction, the fraction of LyC destroyed by dust, non-equilibrium H$\alpha$ emission, and the fraction of H$\alpha$ emission which is due to collisionally ionized and excited hydrogen, however these are in practice not folded into the conversion rate. As for the UV SFRs, we initially use the rates given in \citet{Clarke:2024aa}. For H$\alpha$ this conversion represents stars forming steadily over 10\,Myr with $\mathrm{Z_\ast}=0.004\,\mathrm{Z_\odot}$ and following a \citet{Chabrier:2003aa} IMF with upper mass cutoff of 100\,M$_\odot$, and assuming Case B recombination rates.%,

%\begin{equation}
%\mathrm{SFR_{H\alpha}}=\left( %\frac{\mathrm{L_{H\alpha}}}{\mathrm{erg\,s^{-1}}}\right) \times 10^{-41.59} \,\mathrm{M_\odot\,yr^{-1}}.
%\end{equation}

In Fig.~\ref{fig:sfrint_sfrobs} we show the relationship between SFR$_{10}$ and SFR$_{\mathrm{H}\alpha}$ and the relationship between SFR$_{100}$ and SFR$_{\mathrm{UV}}$. The intrinsic and derived SFRs show a clear correlation, as expected. However, SFR$_{\mathrm{UV}}$ is systematically higher than SFR$_{100}$ and SFR$_{\mathrm{H}\alpha}$ is systematically lower than SFR$_{10}$. Both tracers also show significant scatter with their observationally derived counterparts. 

The scatter for H$\alpha$ is driven largely by the factors already discussed, such as LyC escape and destruction, and by our assumption of a fixed metallicity. Most of these factors, such as LyC escape and destruction, act specifically to decrease the H$\alpha$ flux, and so introduce the systematic offset as well as the scatter.

The UV-derived rates also suffer from scatter partially due to the fixed metallicity assumption, although \citet{Reddy:2022aa} show that this effect is small. The majority of the scatter arises due to the assumption of steady star formation over the last 100\,Myr. In actuality, star formation varies on much shorter timescales, which introduces scatter as well as a bias to higher SFRs. Additionally, the underlying UV emission from older generations of stars creates an SFR floor, which biases the SFRs high. 

We can understand the origin of the scatter with Fig.~\ref{fig:evo}, which shows how the offsets from the SFMS ($\Delta\mathrm{MS}$) change as a function of time. The systematic offsets between the intrinsic and observationally derived SFRs are unimportant for $\Delta\mathrm{MS}$ because the SFMS for each tracer/timescale is fit independently. Here we can see that in general, H$\alpha$ traces SFR$_{10}$ well despite some scatter. However, it is clear that SFR$_\mathrm{UV}$ is tracing SFR$_{100}$ poorly, and is actually probing much shorter timescales.

In order to understand the timescales which are probed by each tracer, we calculate the root mean square deviation \citep[RMSD;][]{Caplar:2019aa, Flores-Velazquez:2021aa, Tacchella:2022aa} between the observationally derived and true SFR as a function of SFR averaging timescale ($t_\mathrm{avg}$)
\begin{equation}
\label{eq:rmsd}
\mathrm{RMSD}(t_\mathrm{avg})= \sqrt{\frac{\sum_i \left( \log\mathrm{SFR}^i_\mathrm{ind} - \log\mathrm{SFR}^i_{t_\mathrm{avg}}\right)^2}{N}}\,,
\end{equation}
where we are summing over each galaxy $i$ of our $N$ galaxies across all snapshots and simulations. $\mathrm{SFR}_\mathrm{ind}$ is the SFR derived by an indicator, whereas $\mathrm{SFR}_{t_\mathrm{avg}}$ is the true SFR at a given averaging timescale. The averaging timescale for which the RMSD is minimized represents the timescale with least scatter, and therefore the timescale best probed by a given tracer. We limit our analysis here to galaxies with SFR$>0.1\,\mathrm{M_\odot yr^{-1}}$ in order to only test the timescale for galaxies where measuring the SFR with H$\alpha$ or UV is most likely to be observationally feasible. In Fig.~\ref{fig:rmsd} we show the normalized RMSD as a function of $t_\mathrm{avg}$. By selecting the range of $t_\mathrm{avg}$ within 1\% of the minimum RMSD, we find that H$\alpha$ best traces star formation on a 5-8\,Myr timescale, whereas UV traces 22--31\,Myr. These results are in broad agreement with the timescales found in other works \citep{Flores-Velazquez:2021aa,Tacchella:2022aa}, although interestingly our UV timescale is closer to the low-redshift than the high-redshift regime reported in \citet{Flores-Velazquez:2021aa}. This highlights that the timescales traced by SFR indicators are dependent on the star-formation histories (SFHs) of galaxies, and therefore on the galaxy formation model assumed.

We provide SFR calibrations suited for high-redshift galaxies with bursty star-formation histories. The advantage of using these calibrations is that they are SFH and stellar metallicity agnostic, and also account for scatter and systematics introduced by complex physics such as LyC escape, which is otherwise difficult to estimate in observed galaxies. We provide the scatter around the relation, which can be added in quadrature to measurement errors to estimate uncertainty due to the SFH, stellar metallicity, and H$\alpha$ physics. We note that this implicitly assumes that the \thzoom simulations span the whole range of possible SFHs, metallicities, and H$\alpha$ physics, which is of course not the case, and therefore there is a level of unaccounted-for systemic bias due to the differences between the \thzoom simulations and reality. While this is a limitation of the method, we still consider this approach advantageous compared to traditional indicators because the state-of-the-art simulations used in this work should, in principle, be closer to reality than the strongly simplified model relied upon for traditional indicators. Based on Fig.~\ref{fig:rmsd}, we select 8\,Myr as the timescale for H$\alpha$ and 24\,Myr as the timescale for UV. Although not strictly the minimum timescale for UV, we deem it more useful to trace longer timescales at the cost of a minor increase in scatter. We calculate our calibration using only star-forming galaxies with SFR$_{8}>0.1\,\mathrm{M_\odot yr^{-1}}$ and SFR$_{24}>0.1\,\mathrm{M_\odot yr^{-1}}$ for H$\alpha$ and UV respectively. Our $\mathrm{SFR_{24}}$ calibration is
\begin{equation}
\label{eq:sfr24}
\mathrm{SFR_{24}}=\left( \frac{\mathrm{\nu L_{\nu,1500}}}{\mathrm{erg\,s^{-1}}}\right) \times 10^{-43.53} \,\mathrm{M_\odot\,yr^{-1}}\,,
\end{equation}
and this relation has a scatter of 0.16\,dex. Our $\mathrm{SFR_{8}}$ calibration is
\begin{equation}
\label{eq:sfr8}
\mathrm{SFR_{8}}=\left( \frac{\mathrm{L_{H\alpha}}}{\mathrm{erg\,s^{-1}}}\right) \times 10^{-41.45} \,\mathrm{M_\odot\,yr^{-1}}\,,
\end{equation}
and this relation has a scatter of 0.14\,dex. 

The impact of using observable-derived SFRs on the SFMS fits is shown in Tab.~\ref{tab:sfms_fits}. The apparent stellar mass dependence is increased for the H$\alpha$ SFMS compared to the intrinsic SFMS, with $\beta=0.077\pm0.015$, whereas the UV SFMS shows a decreased mass dependence. The redshift dependence of the SFMS fit is significantly affected by the use of tracers rather than intrinsic SFRs. The UV SFMS shows a decreased redshift dependence compared to the intrinsic rates, $\mu=2.32\pm0.02$, whereas H$\alpha$ is largely consistent, with $\mu=2.60\pm0.07$.

We note that because we are using the intrinsic UV and H$\alpha$ luminosities, this represents the best-case scenario, where dust can be perfectly corrected for. Dust is expected to have a relatively small contribution for the mostly low-mass and high-redshift galaxies we consider in this work \citep{Maheson:2024aa}.

\begin{figure*} 
\centering
	\includegraphics[width=\textwidth]{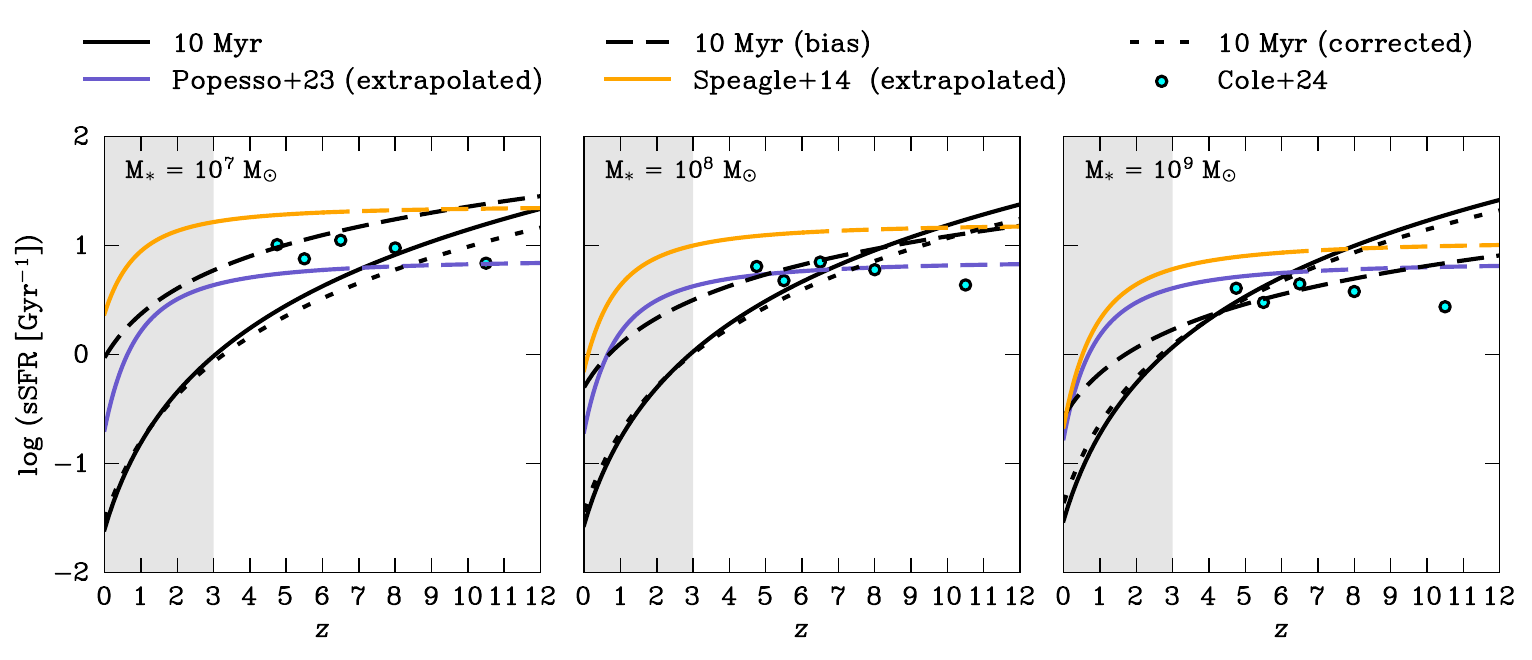}
    \caption{The evolution of sSFR with redshift for galaxies lying on the star-forming main sequence. We show our fit for SFR$_{10}$ (black, Tab.~\ref{tab:sfms_fits}) compared to the SFMS fits of \citet{Speagle:2014aa} (orange line, extrapolated) and \citet{Popesso:2023aa} (blue line, extrapolated), as well as to the redshift-binned fits of \citet{Cole:2025aa} (cyan points). The gray region shows where our fits are extrapolated, and the observational fits are extrapolations where they are dashed. The observational fits diverge at lower masses,  which is mainly because they were fit to more massive galaxies (M$_\ast\,$>$\,10^9\,$M$_\odot$) at $z<6$. Our SFMS shows a much more extreme redshift evolution compared to the observational SFMS fits. We argue that this primarily is due to observational bias, where the slope is artificially flattened due to missing galaxies with low sSFR, i.e. stellar mass incomplete samples. We demonstrate this by fitting the 10\,Myr SFMS excluding galaxies with SFR\,< \,0.1\,M$_\odot\,$yr$^{-1}$. This biased SFMS (black dashed) has a much shallower slope despite the relatively conservative SFR cut. This bias can be corrected by only considering galaxies where sufficiently low sSFRs are still detectable, as seen in our corrected fit where we only consider galaxies with M$_\ast\,$>$\,10^9\,$M$_\odot$ (black dotted). This is well above the typically derived completeness limits for deep \textit{JWST} surveys. }
    \label{fig:sfms_obscomp}
\end{figure*}

\subsection{Comparison with the observed main sequence }
\label{sec:Comparison with literature}

In Fig.~\ref{fig:sfms_obscomp} we show a comparison of the SFMS fit derived in this work to those from the literature. In particular, we show the SFMS from \citet{Speagle:2014aa} and \citet{Popesso:2023aa}, who base their fits on a compilation of other works. Both works are fit on data at $z<6$, and we have extrapolated their fits out to $z=12$ for comparison. The SFMS fits can appear similar when plotted on the SFR versus stellar mass plane because they are dominated by the one-to-one scaling with stellar mass, however the discrepancies become much clearer when considering sSFR. The observational SFMS fits diverge at low masses, which is primarily due to the fact that these fits were based on more massive galaxies (M$_\ast\,$>$\,10^9\,$M$_\odot$).

Our SFMS fits have a significantly steeper slope than the observationally derived fits. Recent \textit{JWST} works have also found weak scaling of the SFMS normalization with redshift \citep{Clarke:2024aa,Cole:2025aa}, indicating that our SFMS is in tension with the fits to observed galaxies. This tension is due to a missing population of low SFR galaxies in observed samples. Our results can be brought into closer agreement with the observed trends by excluding galaxies with SFR\,< \,0.1\,M$_\odot\,$yr$^{-1}$ (dashed; Fig.~\ref{fig:sfms_obscomp}). Due to the trend of increasing SFR with increasing mass and redshift, this cut preferentially removes galaxies with lower masses at lower redshifts. For example, at $z=3$ the remaining M$_\ast\,$=$\,10^8\,$M$_\odot$ galaxies are those at the upper end of the SFMS. This dramatically flattens the redshift evolution and induces a negative dependence of sSFR on stellar mass.

There are two possible causes for the missing low SFR galaxies in observations: stellar mass incomplete samples at high redshift or an overestimation of SFRs in galaxies with intrinsically low SFR. A stellar mass incomplete sample would be likely to miss low SFR galaxies because they tend to be fainter. It has already been noted by \citet{Sun:2023ab} that stellar mass completeness estimates are likely overly optimistic at high redshifts and low masses due to the bursty star formation in these galaxies. Alternatively, low SFR galaxies could be missing from observed samples if SFRs are overestimated for galaxies with intrinsically low SFR. This may be caused by the fact that SFR indicators can also be produced by physical processes other than star formation, such as emission from the older stellar population for UV or collisionally excited H$\alpha$ emission, which may induce a floor in the practically measurable SFR.

This paints a rather pessimistic picture of the prospect for measuring the redshift evolution, but encouragingly if we further restrict our sample to M$_\ast\,$>$\,10^9\,$M$_\odot$ (dotted; Fig.~\ref{fig:sfms_obscomp}), we recover the original trend. By making this further selection we have recovered a nearly stellar mass complete sample because the fraction of galaxies at M$_\ast\,$>$\,10^9\,$M$_\odot$ with SFR\,< \,0.1\,M$_\odot\,$yr$^{-1}$ at $z>3$ is small. In practice, we advise observers to take conservative completeness cuts when fitting the SFMS. A negative slope of sSFR with stellar mass is a warning sign that completeness is having a significant impact on the fit.

We also note that in this section we have investigated the biasing of the intrinsic SFMS. Our UV-derived SFMS fit already shows a shallower redshift dependence than the intrinsic SFMS (see Section~\ref{sec:Observational dependence}). The combination of these observational biases could easily lead to falsely deriving a too-shallow redshift evolution of the SFMS.

\section{Main sequence scatter}
\label{sec:Main sequence scatter}

\begin{figure*}
\centering
	\includegraphics[width=\textwidth]{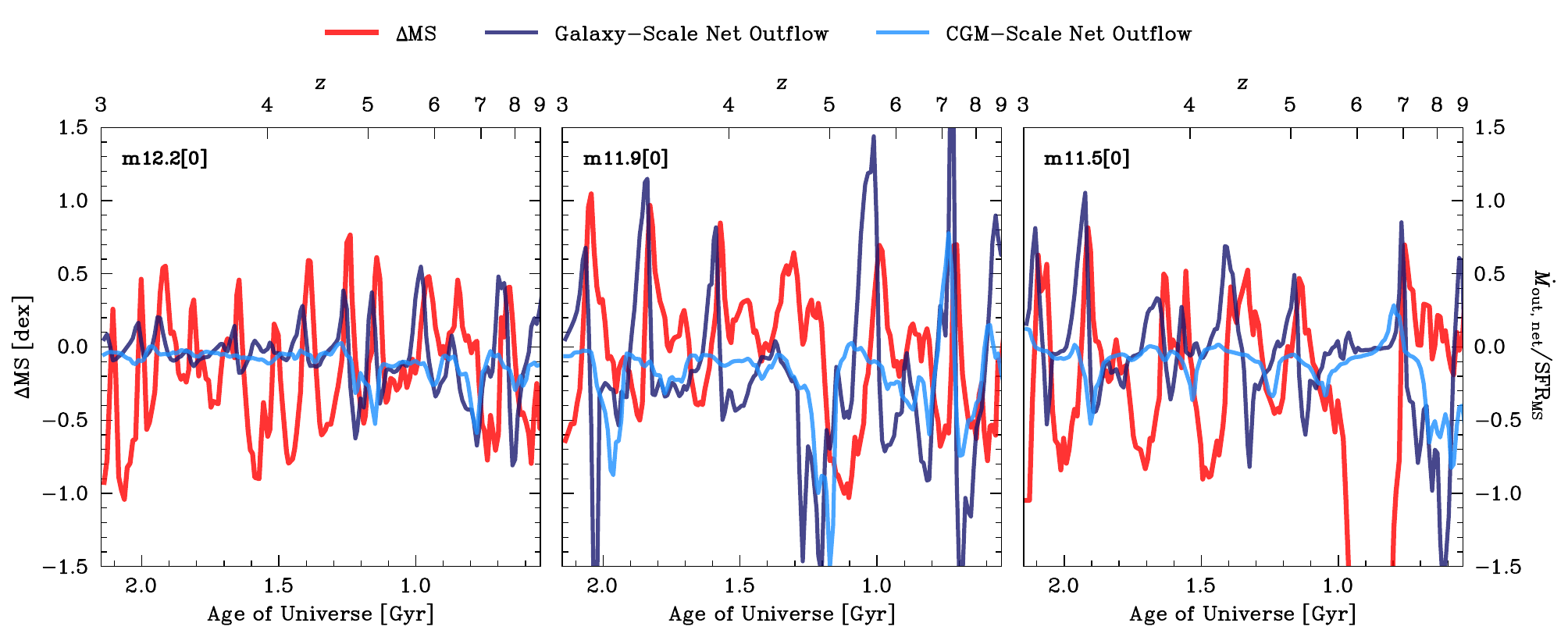}
    \caption{The evolution of the offsets of SFR$_{10}$ from SFMS$_{10}$ ($\Delta\mathrm{MS}_{10}$) and the net galactic/CGM scale net outflow rates, $\dot{M}_\mathrm{out,\,net}$. From left to right we show m12.2 (subhalo 0), m11.9 (subhalo 0), and m11.5 (subhalo 0) with final stellar masses of $10^{10.3}\,\mathrm{M}_\odot$, $10^{10.0}\,\mathrm{M}_\odot$, and $10^{9.5}\,\mathrm{M}_\odot$ respectively. Negative net outflow rates mean that gas is inflowing. We normalize the outflow rates to the SFR corresponding to the SFR for a galaxy on the SFMS at the galaxy's mass and redshift, which contextualizes how significant the inflows/outflows are relative to the typical amount of star formation. When inflow to the CGM is steady, galaxies undergo a breathing mode of star formation where gas cools in the CGM and inflows to the galaxy, leading to a burst of star formation which then generates feedback which causes massive outflows from the galaxy and ceases star formation. However, when there is a significant inflow of gas into the CGM, which can arise due to clumpy inflows or galaxy--galaxy interactions, the starburst is sustained for at least the duration of this inflow.}
    \label{fig:evo2}
\end{figure*}

\subsection{Evolution along the main sequence}
\label{sec:Evolution along the main sequence}

To understand the origin of the main sequence scatter, we consider how galaxies evolve along the main sequence. In particular, we are interested in how SFMS offsets are related to gas inflow and outflow rates. We consider the outflow rates at two radii; twice the UV half-light radius as a proxy for the galaxy or ISM scale, and the $0.7R_\mathrm{vir}$ as a proxy for the CGM scale. This allows us to distinguish the exchange of gas between the CGM and ISM from the accretion of gas to the halo from the IGM and mergers. To calculate the outflow rates, we follow \citet{Nelson:2019aa},
\begin{equation}
\dot{M}_{\mathrm{gas}}= {\frac{\delta M_\mathrm{gas}}{\delta t}\,\vrule\,}_\mathrm{rad} =\frac{1}{\Delta r} \sum_{\substack{i=0 \\ \vert r_i - r_0\rvert<\Delta r/2}}^{N} \left( \frac{\mathbf{v_i \cdot r_i}}{\vert r_i \rvert} m_i \right)\,,
\end{equation}
where we are summing over all gas cells $i$ within a shell of width $\Delta r$ centered around $r_0$ with $\Delta r = 0.1 r_0$. In general, unless otherwise stated, we consider the net inflow/outflow rates, and so we consider all particles. 

\begin{figure*} 
\centering
	\includegraphics[width=0.85\textwidth]{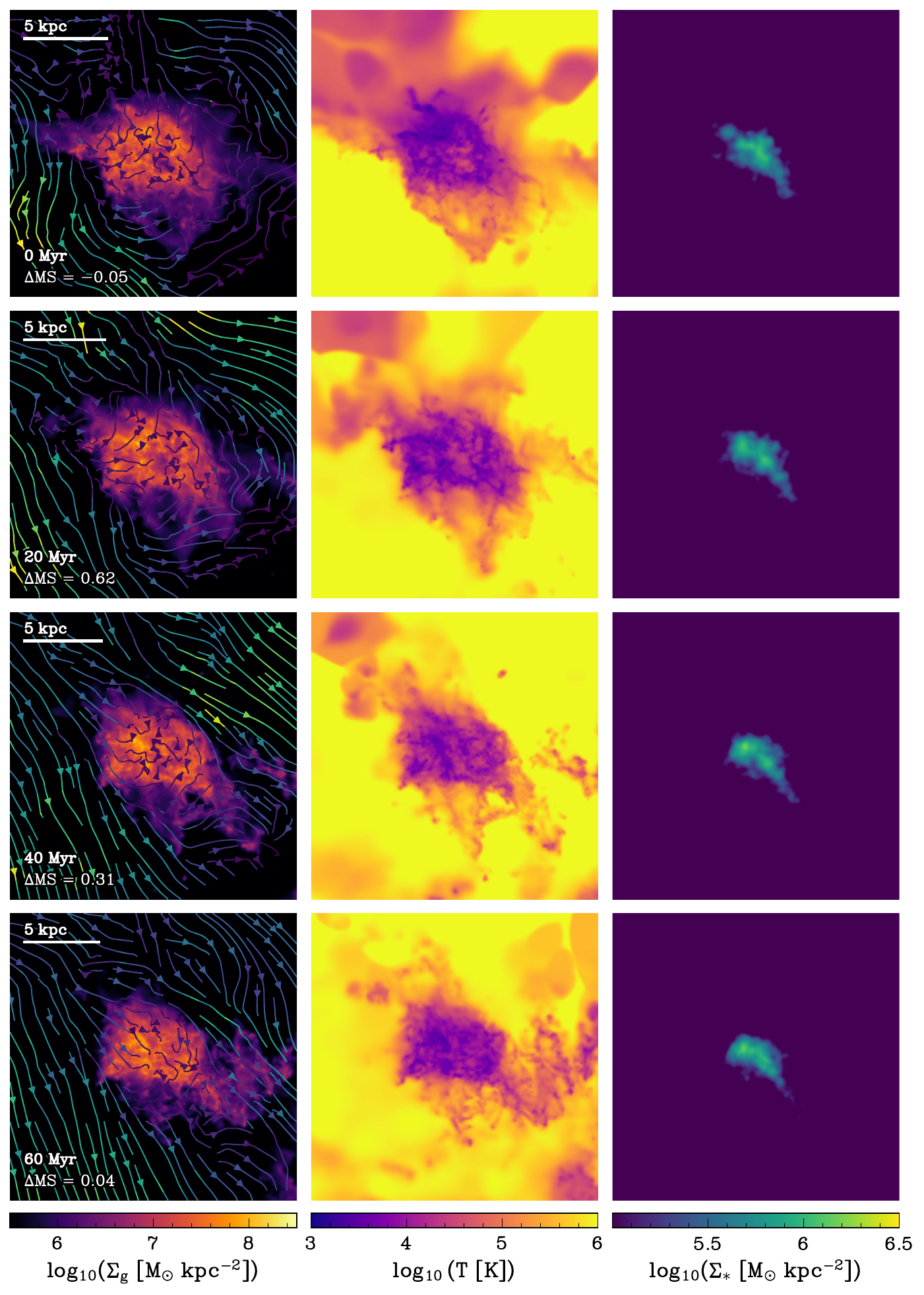}
    \caption{A $M_\ast=10^{6.9}\,\mathrm{M_\odot}$ galaxy (m11.1, subhalo 2) undergoing an internally driven burst phase where inflow to the CGM is constant throughout. A 60\,Myr timeline is shown from top to bottom, beginning at $z=4.2$. We show the gas surface density overlaid with velocity streamlines colored by the velocity magnitude (left), mass-weighted gas temperature (middle), and stellar surface density (right). The projection has dimensions $2R_\mathrm{vir}\times2R_\mathrm{vir}\times10\,\mathrm{kpc}$. The starburst drives outflows on the ISM scale, quenching star formation, however there are no massive outflows on the CGM scale. Steady accretion to the CGM is largely uninterrupted.
    }
    \label{fig:internal_gas_image}
\end{figure*}

Fig.~\ref{fig:evo2} shows the offset from the 10\,Myr averaged SFMS, $\Delta\mathrm{MS}_{10}$, alongside the galaxy and CGM scale net outflow rates. Qualitatively, we can see that bursts are generally preceded by an inflow of gas on the galaxy scale, and shortly followed by an outflow of gas from the galaxy to the CGM. This burst-quench cycle appears to proceed even when when inflows on the CGM scale are smooth with time. This mode of burstiness appears to be driven by processes internal to the galaxy and resembles the ``breathing'' mode of star formation discussed in \citet{Hopkins:2023aa}, where star formation ceases rapidly due to the shallow gravitational potentials being unable to retain the ISM. In Fig.~\ref{fig:internal_gas_image}, we show an example of an internally-driven starburst. The CGM inflow rate remains steady throughout the burst and in the galaxy's recent history. The burst is short lived, with the galaxy spending only $\sim$20\,Myr above the main sequence. While there are galaxy-scale outflows, there is little impact of this starburst on a larger scale, with no visible CGM-scale outflows.

\begin{figure*} 
\centering
	\includegraphics[width=0.85\textwidth]{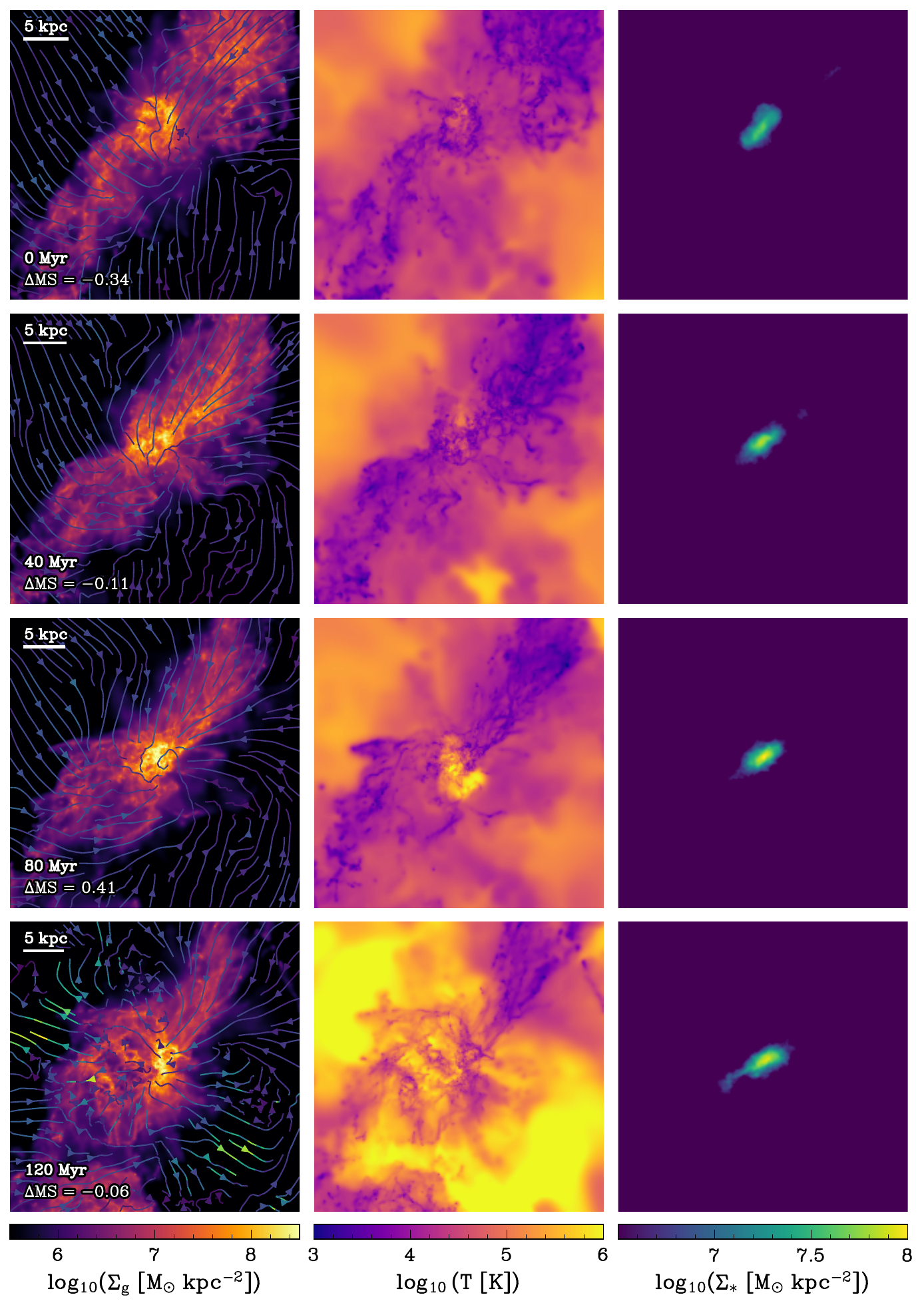}
    \caption{Same as Fig.~\ref{fig:internal_gas_image} but for a $M_\ast=10^{8.5}\,\mathrm{M_\odot}$ galaxy (m11.5, subhalo 0) undergoing an externally driven burst phase, in this case due to a merger. A 120\,Myr timeline is shown from top to bottom, beginning at $z=5.2$. Massive outflows are driven from the galactic scale, evacuating the ISM and quenching star formation. In the less dense regions of the CGM, fast-moving hot outflows expel gas to the IGM, whereas in more dense filamentary regions, gas accretion to the ISM is interrupted.}
    \label{fig:external_gas_image}
\end{figure*}

However, in Fig.~\ref{fig:evo2} we also see that the inflow onto the CGM is often clumpy, and that there can be extended periods of significant gas accretion. The inflow clumpiness can be due to clumpy inflows of IGM gas, or due to gas accreted in galaxy--galaxy interactions. The clumpiness of gas inflow may be aided by periods of inflow stream counter rotation relative to the halo angular momentum \citep{Tacchella:2016aa}. Following the initiation of a period of heightened CGM gas accretion, a burst of star formation begins and lasts for at least as long as the heightened halo gas accretion. The length of these bursts can be significantly longer-lasting than those of the internally driven mode, and therefore does appear to be a second, distinct mode of bursty star formation which is externally driven. In this mode, massive inflows to the CGM cool and provide significant fuel for star formation. Inflows come in as cold streams or via cold gas through mergers, allowing star formation to continue while feedback is vented out via the lower-density regions of the CGM. In Fig.~\ref{fig:external_gas_image}, we show an example of an externally-driven starburst. A gas-rich wet double merger provides a significant inflow of gas to the galaxy, triggering a long-lasting and extreme burst of star formation. Massive, hot CGM scale outflows are visible following the burst, mainly proceeding along the lower-density regions of the CGM. 

\begin{figure*} 
\centering
	\includegraphics[width=\textwidth]{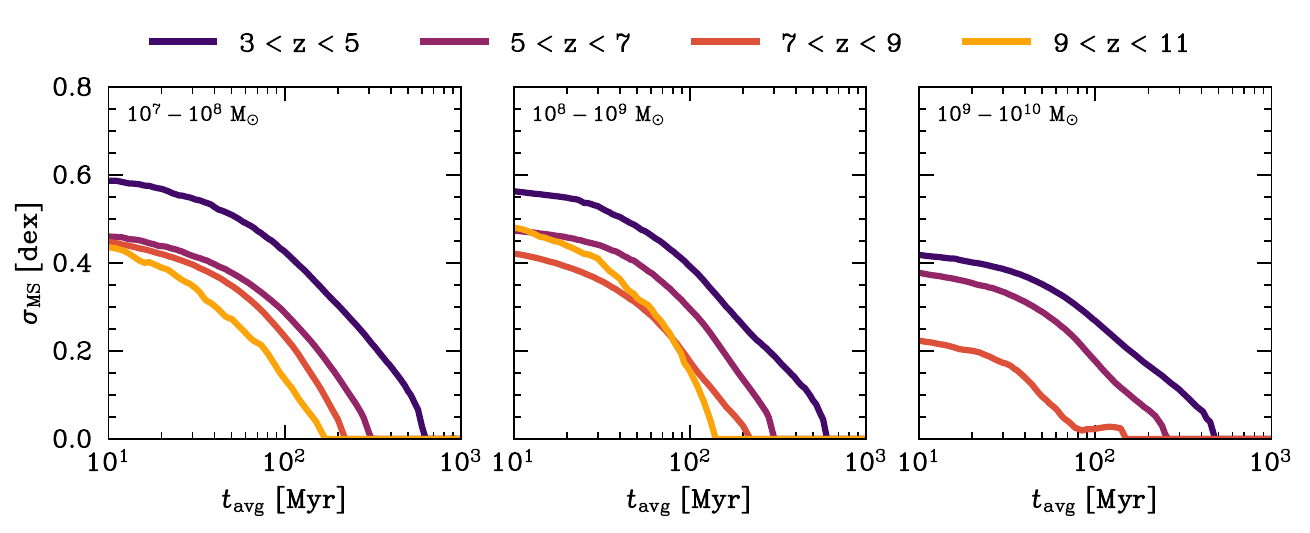}
    \caption{The scatter of the star-forming main sequence ($\sigma_\mathrm{MS}$) as a function of the averaging timescale ($t_\mathrm{avg}$) for three stellar mass bins. We only show $z<9$ for the highest mass bin due to insufficient sample size. The scatter approaches 0\,dex with increasing $t_\mathrm{avg}$ as multiple bursts are averaged over, and must reach 0 at the age of the Universe for each redshift as the stellar mass formed within the averaging time and the total stellar mass become equivalent. The scatter increases with cosmic time (decreasing redshift) due to the importance of long-term environmental effects increasing. The scatter also decreases at higher stellar mass, partially due to the reduced impact of the environment on massive galaxies, which are unlikely to be satellites, but also potentially indicating the emergence of steady star formation in massive galaxies.}
    \label{fig:sfms_scat}
\end{figure*}

\subsection{Connecting the burst--quench cycle to outflows}
\label{sec:The relationship between the burst-quench cycle and outflows}

To quantitatively understand the relationship between star formation and outflow rates on the galaxy and halo scale, we perform a cross-correlation analysis. For each merger tree in our halo catalogue, we follow the main (most massive) progenitor through snapshots to create a time series of $\mathrm{SFR}_{10}$ and the net inflow rate. The snapshots are not evenly spaced in time so we linearly interpolate to a 1\,Myr grid. 

We perform the cross-correlation separately on net inflow and net outflow rates. This is done by taking the net outflow rate and setting the value to 0 for values greater (less) than 0 when we are performing cross-correlation on the inflow (outflow) rates. We also restrict the fitting such that inflow rates can only lag before $\mathrm{SFR}_{10}$ and so that outflow rates can only lag behind. This prevents inflow/outflow events from being associated with an unrelated star-formation event. We then cross-correlate each quantity with $\mathrm{SFR}_{10}$ for each subhalo to find the lag-time with the highest correlation. To ensure there is sufficient data to perform the analysis, we require that the subhalo has at least 100 data points, corresponding to a 100\,Myr history. We then generate summary statistics such as the median lag time and correlation coefficient.

We find strong correlations on the galactic scale. For $\mathrm{SFR}_{10}$ and galactic net outflow rates, across all galaxies in our sample we find a median correlation of $r=0.70\pm0.18$ and a median lag time of $t_\mathrm{lag}=21${\raisebox{0.5ex}{\tiny$^{+22}_{-13}$}}\,Myr. For $\mathrm{SFR}_{10}$ and galactic net inflow rates, we find a median correlation of $r=0.61\pm0.24$ and a median lag time of $t_\mathrm{lag}=-41${\raisebox{0.5ex}{\tiny$^{+19}_{-33}$}}\,Myr (the negative value of $t_\mathrm{lag}$ indicates that the inflow precedes the starburst, as expected). The errors on these quantities reflect the $16^\text{th}$ and $84^\text{th}$ percentiles. In general, we can see that starbursts appear to be strongly related to recent inflows to the ISM and followed by strong, ISM-clearing outflows.

On the CGM scale, the correlations are weaker. This reflects the fact that not all starburst events are driven by a recent period of heightened inflow. For $\mathrm{SFR}_{10}$ and CGM net outflow rates, we find a median correlation of $r=0.44\pm0.34$ and a median lag time of $t_\mathrm{lag}=101${\raisebox{0.5ex}{\tiny$^{+64}_{-66}$}}\,Myr. For $\mathrm{SFR}_{10}$ and CGM inflow rates, we find a median correlation of $r=0.45\pm0.30$ and a median lag time of $t_\mathrm{lag}=-106${\raisebox{0.5ex}{\tiny$^{+91}_{-245}$}}\,Myr. The correlation coefficient for the CGM scale inflow and outflow rates are close in value, possibly reflecting the fact that net outflows from the CGM scale are only triggered by massive starbursts, which themselves are generally driven by a recent massive inflow on the CGM scale. 

\begin{figure} 
\centering
	\includegraphics[width=\columnwidth]{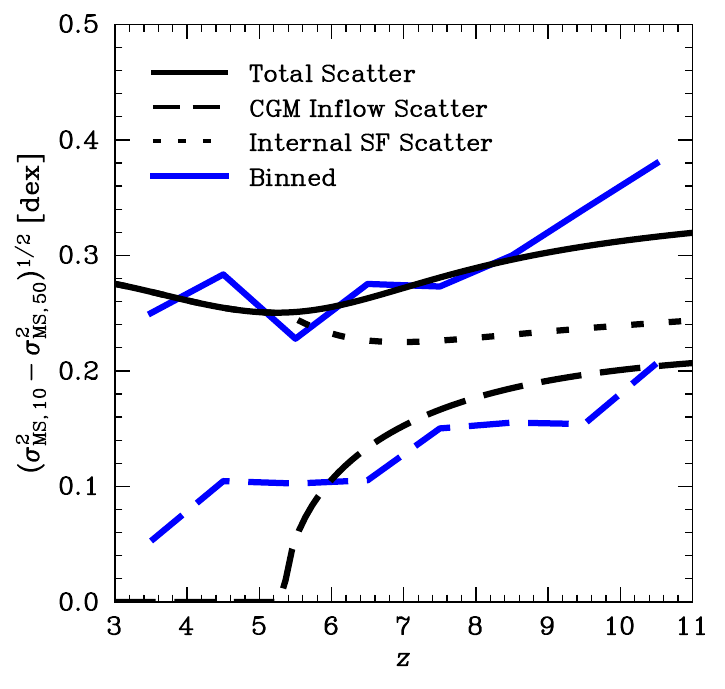}
    \caption{Short-term burstiness, characterized as $(\sigma_\mathrm{MS,10}^2-\sigma_\mathrm{MS,50}^2)^{1/2}$, as a function of redshift. SFMS scatter (solid), CGM inflow scatter (dashed), and scatter due to internal processes (dotted) are shown. Quantities are shown for the broken power-law fit to $\sigma_\mathrm{MS}$ (Eq.~\ref{eq:theta}, calculated at $\mathrm{M}_{\ast} = 10^{8}\,\mathrm{M}_{\odot}$, black) and for the scatter of $10^7\,\mathrm{M}_{\odot} < \mathrm{M}_{\ast} < 10^{10}\,\mathrm{M}_{\odot}$ galaxies measured in redshift bins (blue). The scatter in the SFMS on short timescales increases with redshift, despite the total scatter decreasing (see Fig.~\ref{fig:vir_inflow_ratios}). Internal processes are responsible for the bulk of short-term SFMS scatter, however they show weak redshift evolution. The increase in short-term SFMS scatter with redshift is due to the increase in CGM inflow scatter. This implies that while galaxy burstiness overall is dominated by internal processes, much of the increase in galaxy burstiness with redshift is due to more rapidly varying gas inflow rates. }
    \label{fig:feld_ratios}
\end{figure}

\begin{figure} 
\centering
	\includegraphics[width=\columnwidth]{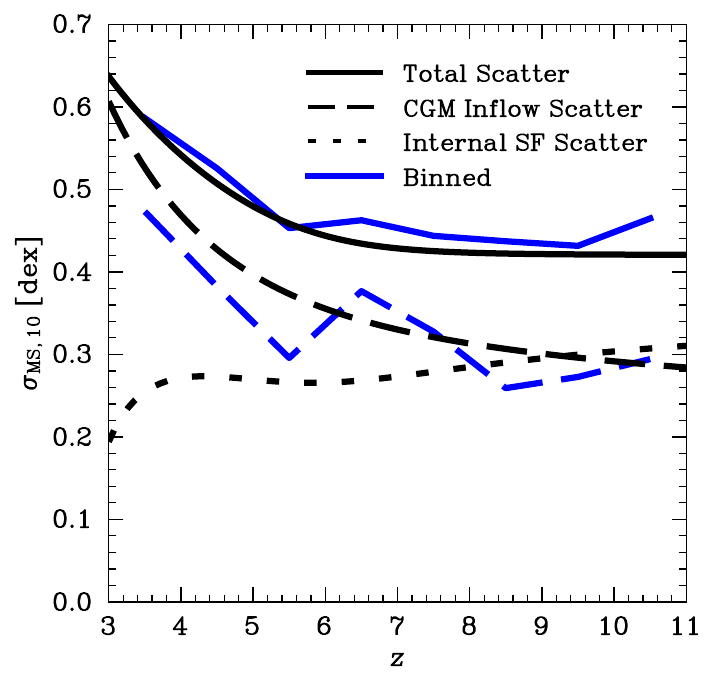}
    \caption{The scatter of the SFMS (solid) declines with redshift,  which is accounted for by reduced scatter in the CGM inflow main sequence (dashed). Internal processes (dotted) provide a baseline of scatter, which cause the SFMS scatter to approach a constant value at high redshift. Quantities are shown for the broken power-law fit to $\sigma_\mathrm{MS}$ (Eq.~\ref{eq:theta}, black) and for the scatter of $10^7\,\mathrm{M}_{\odot} < \mathrm{M}_{\ast} < 10^{10}\,\mathrm{M}_{\odot}$ galaxies measured in redshift bins (blue). Inflow scatter dominates at low redshift due to longer-term environmental effects such as starvation.}
    \label{fig:vir_inflow_ratios}
\end{figure}

In summary, we find that galactic scale inflows and outflows are strongly correlated with the burst-quench cycle, and that the time delay between galactic scale inflows, outflows, and starbursts is short. CGM-scale inflows and outflows are more weakly correlated with starburst events and over longer timescales, demonstrating that rapid gas inflow can cause massive starbursts which in turn drive CGM-scale outflows, but that rapid CGM-scale inflows are not a requirement for bursty star formation in general. We note that \citet{Saldana-Lopez:2025aa} recently found a strong association between outflows and bursty star formation in galaxies observed during the epoch of reionization using observations with \textit{JWST}, which is consistent with our findings here.

\subsection{Evaluating main sequence scatter}
\label{sec:Evaluating main sequence scatter}

The scatter of the SFMS for SFRs of different averaging timescales ($t_\mathrm{avg}$) encodes information about the variability of star formation. We are interested in how this scatter evolves as a function of mass and redshift, so we split our galaxies in redshift and stellar mass bins. We calculate the offsets from the SFMS ($\Delta\mathrm{MS}$) of each timescale for all galaxies, which allows us to remove redshift and mass-dependence induced scatter within each bin. For each bin and timescale, we fit a zNDB distribution and define the SFMS scatter ($\sigma_\mathrm{MS}$) following the symmetric scatter approximation \citep{Feldmann:2017aa}. This scatter is analogous to the scatter of a log-normal distribution. For completeness, we show other, less accurate measures of the scatter in Appendix~\ref{sec:Different measures of scatter}.

In Fig.~\ref{fig:sfms_scat} we show $\sigma_\mathrm{MS}$ as a function of SFR averaging timescale for each redshift and stellar mass bin. For each bin, the $\sigma_\mathrm{MS}$ decreases with increasing $t_\mathrm{avg}$. This is expected because the scatter at a given timescale is the combination of scatter caused at that exact timescale and all longer timescales. $\sigma_\mathrm{MS}$ approaches zero as $t_\mathrm{avg}$ increases and multiple bursts are averaged over. $\sigma_\mathrm{MS}$ must reach zero by the age of the Universe at a given redshift because all stars in a galaxy must have been formed within the age of the Universe essentially making stellar mass and $\mathrm{SFR}\times t_\mathrm{avg}$ equivalent.

$\sigma_\mathrm{MS}$ decreases with redshift across all averaging timescales. This goes against the naive expectation that $\sigma_\mathrm{MS}$ would increase with redshift due to the idea that galaxies are more bursty in the early Universe. However, it is important to note that $\sigma_\mathrm{MS}$ at a given timescale encodes the scatter from all longer timescales; at lower redshifts, much of the scatter is due to longer-timescale effects, such as environmental influence. Our interpretation of environmental influence is consistent with the reduced redshift evolution of the scatter when satellites are excluded (see Appendix~\ref{sec:Impact of centrals and satellites}).

We can isolate the scatter caused by shorter-term processes by considering the increase in scatter on short timescales using the quantity $(\sigma_\mathrm{MS,10}^2-\sigma_\mathrm{MS,50}^2)^{1/2}$. In Fig.~\ref{fig:feld_ratios} we show that scatter on short timescales (<50\,Myr) does indeed increase with redshift. In addition to the scatter calculated in mass bins, we also show the best fit broken power law scatter derived during the SFMS fits (see Section~\ref{sec:Measuring the star-forming main sequence}) for a galaxy at M$_\ast=10^8~\mathrm{M}_\odot$. These fits have the advantage of being less prone to noise because they use all of the data, but the parametric nature means they are not able to encode all relevant behavior.

In Section~\ref{sec:Evolution along the main sequence} we introduced externally and internally driven star-formation bursts. Using the SFMS scatter, we can understand the characteristics and importance of these two modes. To calculate the scatter due to the external mode ($\sigma_\mathrm{MS,ext}$), which is driven by variability of CGM-scale inflows, we fit a ``CGM-inflow main sequence'', for which we follow the same procedure as for the SFMS except we use the CGM-scale inflow rate instead of the SFR. We find the scatter in the CGM-inflow main sequence in the same manner as for the SFMS, and define this as $\sigma_\mathrm{MS,ext}$. The remaining scatter is due to internal processes, and we calculate this as $\sigma_\mathrm{MS,int}=(\sigma_\mathrm{MS}^2-\sigma_\mathrm{MS,ext}^2)^{1/2}$. We note that this approach means that any scatter that is not due to CGM-scale inflows are defined as part of the internal scatter.

In Fig.~\ref{fig:vir_inflow_ratios} we show $\sigma_\mathrm{MS,ext}$ and $\sigma_\mathrm{MS,int}$ as a function of redshift for $t_\mathrm{avg}=10\,\mathrm{Myr}$. $\sigma_\mathrm{MS}$ and $\sigma_\mathrm{MS,ext}$ decrease with redshift, reflecting the previously mentioned longer-term environmental effects becoming less important at higher redshifts. Other simulations with on-the-fly radiative transfer have similarly found a dramatic impact on the inflow of gas to low-mass haloes during reionisation \citep[e.g.,][]{Katz:2020aa}. $\sigma_\mathrm{MS,int}$ remains roughly constant with redshift, showing that the internal mode of star-formation bursts is primarily an effect of mass. $\sigma_\mathrm{MS,ext}$ accounts for the bulk of the scatter in $\sigma_\mathrm{MS}$, except for the highest redshifts. This demonstrates that star-formation variability is mostly dominated by variability in CGM-scale inflows, except for the highest redshifts, where the scatter in inflow rates approaches the constant scatter floor which is set by physics internal to the galaxy.

If the scatter is mostly driven by CGM scale inflow variability, what causes the apparent increase in burstiness with redshift? We return to Fig.~\ref{fig:feld_ratios} to investigate this, where we show $(\sigma_\mathrm{MS,10}^2-\sigma_\mathrm{MS,50}^2)^{1/2}$ for both the internal and external scatter. The internal scatter on short timescales shows little evolution with redshift, again demonstrating that the internal mode is mostly dependent on stellar mass. However, the external scatter on short timescales increases with redshift. Externally driven starbursts are more common at high redshift, and this is the main cause of galaxies being more bursty on short timescales in the early Universe. The increased prevalence of externally driven starbursts could be caused by several reasons, including more clumpy CGM-scale inflows and stochastic gas supply via mergers. The feedback from galaxies themselves likely contributes somewhat to the clumpiness of the gas supply at high redshift. sSFRs are higher at high redshift, and therefore the starburst events are more intense. The feedback from these more intense starbursts can halt the accretion of gas from the IGM, as seen in Fig.~\ref{fig:external_gas_image}, and therefore sufficiently strong previous starburst events can increase the clumpiness of the CGM inflow in the galaxy's future. We also note that the dynamical times of haloes are shorter in the early Universe, which may naturally shift the characteristic inflow timescale to shorter times.

\begin{figure*} 
\centering
	\includegraphics[width=\textwidth]{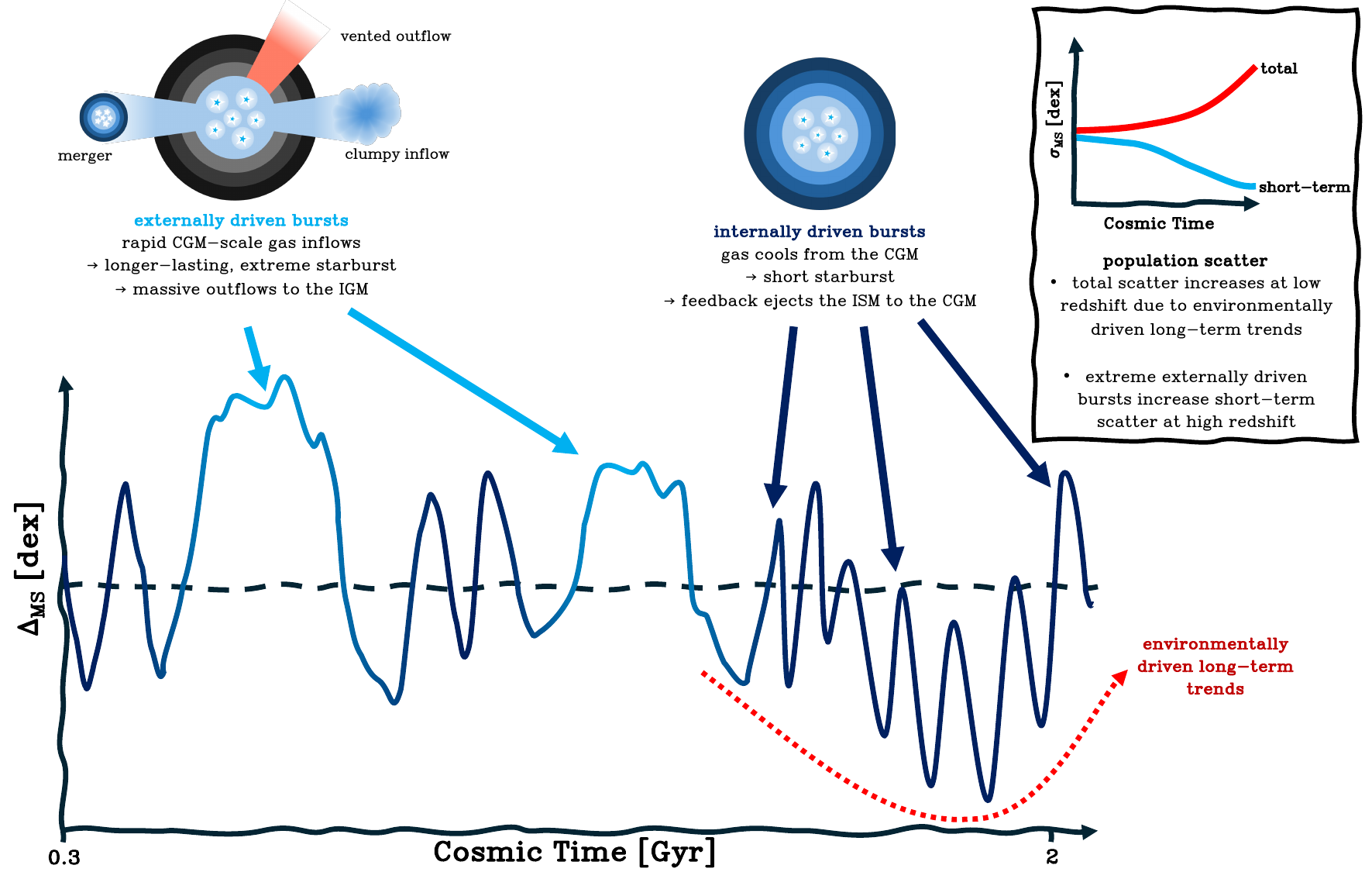}
    \caption{Diagram showing how an individual galaxy's evolution along the SFMS, shown here with the offset from the SFMS ($\Delta_\mathrm{MS}$), can give rise to SFMS scatter ($\sigma_\mathrm{MS}$) trends for a population of galaxies (top right). Externally driven bursts, which are triggered by the rapid and efficient delivery of cold gas from the IGM to the galaxy, are more common at early times due to the increased short-term variability of inflows to the CGM. These bursts are long-lasting periods of intense star formation, and so the short-term scatter in the SFMS is largest at early times. Internally driven bursts, where burstiness is caused by the ejection of the ISM due to strong stellar feedback and a shallow gravitational potential, are primarily a mass-dependent effect which can happen across cosmic time and therefore set a baseline in the short-term SFMS scatter. At later times, environmental effects can restrict the inflow of gas to the CGM, long-term trends in $\Delta_\mathrm{MS}$ as the galaxy is starved of star-formation fuel. Across an ensemble of galaxies, this leads to an increase in the total main-sequence scatter.}
    \label{fig:ms_scheme}
\end{figure*}

We illustrate the arguments made in this section in Fig.~\ref{fig:ms_scheme}, where we sketch out how the evolution of individual galaxies can lead to the trends in $\sigma_\mathrm{MS}$. Highly variable CGM inflows at early times lead to an increase in externally driven bursts, which are so extreme on short timescales that they lead to maximum short-term scatter at high redshift. Internal bursts can occur across cosmic time, and set a baseline for the short-term SFMS scatter. Long-term trends in SFMS offsets become apparent at low redshift, driven by environmental effects such as the suppression of gas inflow to the CGM. This leads to the overall scatter of the SFMS being highest at low redshifts. Although our simulations terminate at $z=3$, we speculate that the scatter would further increase down to $z=0$ as the diversity of environments increases. We note that because the \thzoom simulations represent a biased sampling of cosmological volumes which may affect our results. In particular,  low-mass galaxies included in our sample are likely located closer to massive galaxies on average than a typical low-mass galaxy in reality. Future work using larger volume simulations will be needed to clarify these effects.

\subsection{Characterizing burst and quenching events}
\label{sec:Characterizing burst and quenching events}

\begin{figure} 
\centering
	\includegraphics[width=\columnwidth]{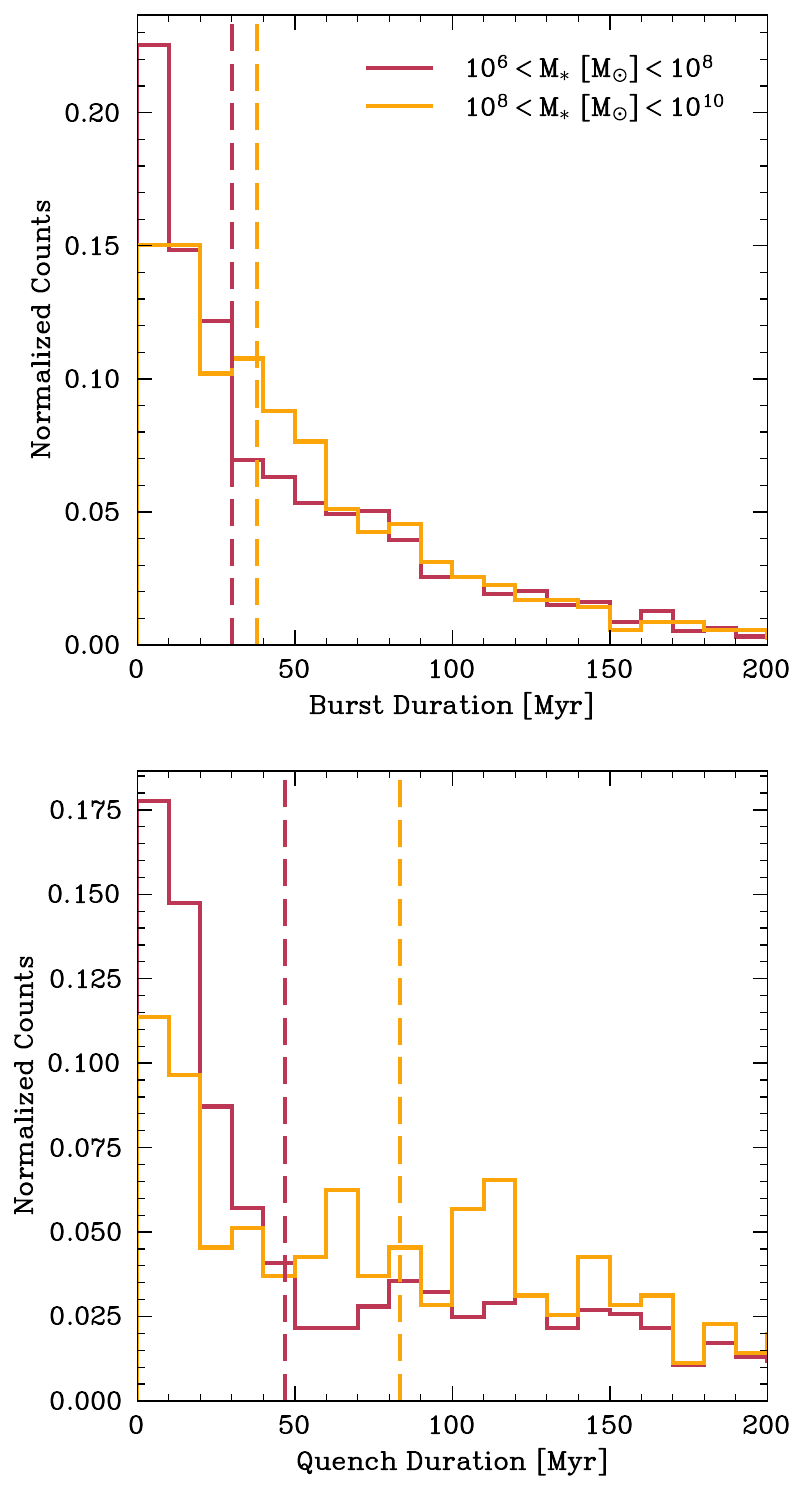}
    \caption{The distribution of burst and quench event durations. The dashed vertical lines show the median duration of a stellar mass bin. \textit{Top:} The distribution of burst durations for two stellar mass bins. We define a burst as the time spent above the main sequence, and a quench event as the times spent below the main sequence, which is appropriate for the bursty star formation we consider. \textit{Bottom:} The duration of quenching events for two stellar mass bins. The duration of burst and quench events increases with stellar mass. Quenching events generally last longer than bursts due to the self-regulating nature of starbursts. A galaxy can remain quenched for an arbitrarily long amount of time, but a starburst will deplete and eject the ISM, causing star formation to cease.}
    \label{fig:burst_length_main_histogram}
\end{figure}

We can use offsets from the SFMS to investigate the nature of burst and quench events. In Fig.~\ref{fig:burst_length_main_histogram}, we show the durations of burst and quench events. We define a burst as the period of time when $\Delta\mathrm{MS}_{10}>0$ and a quench event as when $\Delta\mathrm{MS}_{10}<0$. This definition is sufficient for the galaxies we consider, which are undergoing burst-quench cycles.  We consider two stellar mass bins, $10^6\,\mathrm{M}_{\odot} < \mathrm{M}_{\ast} < 10^{8}\,\mathrm{M}_{\odot}$ and $10^8\,\mathrm{M}_{\odot} < \mathrm{M}_{\ast} < 10^{10}\,\mathrm{M}_{\odot}$, which have median burst durations of 30 and 38\,Myr, respectively, and median quench durations of 47 and 83\,Myr, respectively.

In general, quenching events can last longer than burst events at the same stellar mass. This likely reflects the fact that a galaxy can quench completely due to a lack of inflows and remain quenched for an arbitrary length of time, whereas starbursts in these bursty galaxies are self-regulating; i.e., a galaxy which forms enough stars to rise far above the SFMS will also generate enough feedback to cause star formation to cease. The conversion of gas into stars also depletes the gas reservoir, further limiting the extremity of starbursts. Analyzing mass-loading factors in detail is beyond the scope of this work, but initial analysis indicates that this conversion can be an important factor in depleting the ISM during starbursts.

We investigated but were unable to identify any significant trend with redshift. This is likely due to our small sample size; high redshifts make up a small fraction of cosmic time, so our sample of burst and quenching events is small even though we have many galaxies. Our earlier analysis of the SFMS leads us to expect that the durations of bursts and quenching events should also show significant evolution with redshift if our sample were large enough.

\subsection{UV scatter}
\label{sec:UV scatter}

\begin{figure} 
\centering
	\includegraphics[width=\columnwidth]{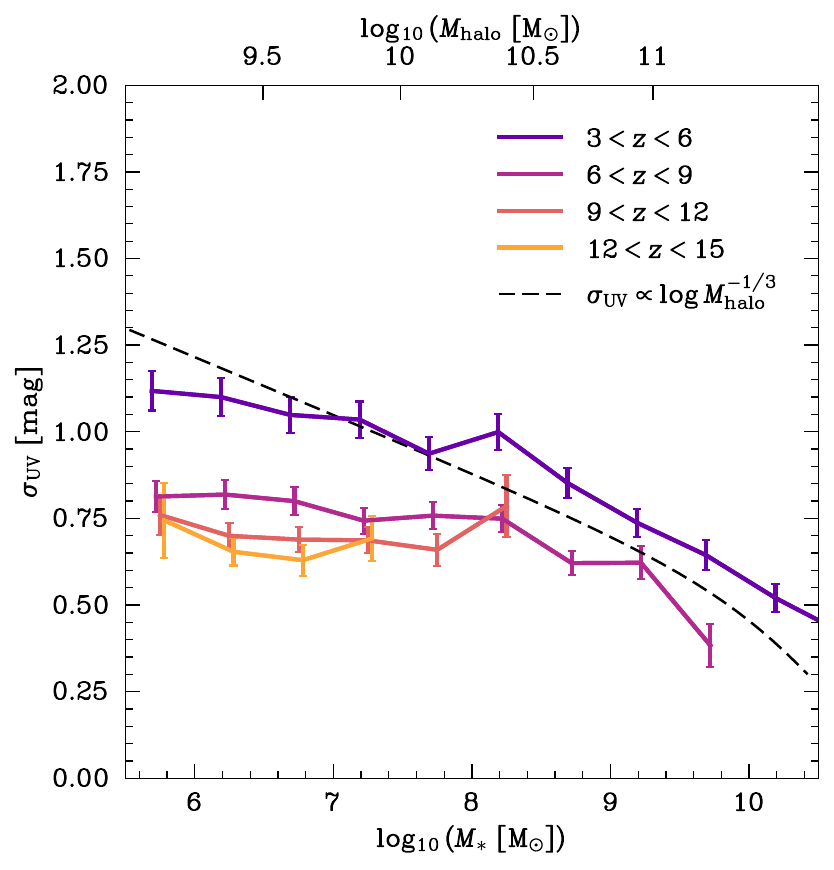}
    \caption{Intrinsic UV (1500\AA) scatter, $\sigma_\mathrm{UV}$, as a function of stellar mass for four redshift bins. The top y-axis shows the halo mass following the $M_\ast$--$M_{\mathrm{halo}}$ relation presented in \citet{Tacchella:2018aa}. Scatter due to dust extinction is unaccounted for here, however there is no increase in scatter when using UV emission attenuated with the fiducial \thzoom dust model. At $z>6$, the scatter converges to $\sigma_\mathrm{UV}\approx0.7\,\mathrm{mag}$. This indicates that increasing UV scatter at high redshift alone may not explain the over-abundance of bright galaxies seen with \textit{JWST}. The black dashed line shows the analytic scatter following \citet{Gelli:2024aa}, which was shown by \citet{Shen:2025aa} to reproduce \textit{JWST}-derived UVLFs when combined with the \thzoom star-formation efficiency. While the scatter shown here agrees with this trend at $M_\ast\gtrsim 10^{8}\mathrm{M_\odot}$, our scatter flattens off at lower masses. This implies that additional scatter is needed at lower masses, or that additional UV luminosity is needed in lower mass galaxies, perhaps due to a top-heavy IMF or UV emission from AGN.}
    \label{fig:uv_scatter}
\end{figure}

In this section, we relate our models to the apparent over-abundance UV-bright galaxies observed with \textit{JWST} \citep[e.g.][]{Donnan:2023aa,Donnan:2024aa,Leethochawalit:2023aa,Perez-Gonzalez:2023aa,Robertson:2024aa,Whitler:2025aa}, and in particular to whether bursty star-formation can naturally explain them \citep[e.g.][]{Shen:2023aa,Sun:2023aa,Kravtsov:2024aa}. In Fig.~\ref{fig:uv_scatter} we show the intrinsic UV scatter, $\sigma_\mathrm{UV}$, as a function of stellar mass. We calculate the scatter using the intrinsic UV emission, however calculation with the observed UV emission does not yield increased scatter because the dust in the \thzoom simulation is generally weak (Garaldi et al. in prep.). We also note that when considering $\sigma_\mathrm{UV}$ as a function of halo mass, there is additional scatter between the halo mass and stellar mass not considered here.

In calculating $\sigma_\mathrm{UV}$, we have also loosened our selection of galaxies such that we require only a minimum of 100 stellar particles and we do not require the galaxy to be traced from a valid $z=3$ subhalo. We found this selection acceptable because older stars also contribute to the UV emission, which causes UV magnitude distributions to be more Gaussian than star-formation rate distributions. This means that the UV scatter is less sensitive to the sSFR cutoff than the SFMS scatter. The more lax sample selection causes a negligible change in the scatter calculated at $z<12$, but it importantly increases the sample size enough to allow us to calculate the scatter at $12<z<15$.

At $z>6$, we find that the scatter converges to $\sigma_\mathrm{UV}\approx0.7\,\mathrm{mag}$ for galaxies at $M_\ast\lesssim 10^{8}\mathrm{M_\odot}$. Although galaxies are becoming more bursty as they have increased variability on short timescales, their overall scatter is not increasing because longer-term scatter is only introduced with increasing cosmic time (see Section~\ref{sec:Evaluating main sequence scatter}).  While we do not consider the additional scatter which may arise due to dust attenuation, we do not expect this to increase strongly with redshift. 

Overall this leads us to a picture where although our galaxies are becoming more bursty at high redshift, this does not lead to an increase in $\sigma_\mathrm{UV}$ independent of mass. This implies that we do not find the increasing scatter out to $z=15$ that may be needed to explain \textit{JWST} luminosity functions \citep{Shen:2023aa,Shen:2024ab,Kravtsov:2024aa,Gelli:2024aa,Kravtsov:2024aa}. Additionally, although we find a trend of higher scatter at lower masses for galaxies with $M_\ast\gtrsim 10^{8}\mathrm{M_\odot}$, this trend flattens off at lower masses. This is an issue because a trend of increasing scatter with decreasing stellar mass combined with the increased prevalence of low-mass galaxies at high redshift may be sufficient to effectively increase the scatter at the population level \citep{Gelli:2024aa,Shen:2024ab}. For example, \citet{Shen:2025aa} show that the star-formation efficiency in the \thzoom simulations can reproduce the UV luminosity function out to $z=14$ following the \citet{Gelli:2024aa} halo mass-dependent scatter, $\sigma_{\mathrm{UV}} \propto \log M_{\mathrm{halo}}^{-1/3}$. Additional scatter may be introduced at lower halo masses due to scatter in the $M_\ast$--$M_{\mathrm{halo}}$ relation, although typically this is found to be around 0.2--0.3\,dex, which would not be sufficient \citep{Tacchella:2018aa,Shen:2023aa}. Therefore, it seems that at low stellar and halo masses, either additional scatter or an increase in UV luminosity is required. This could be achieved through various means, such as a top-heavy IMF \citep{Shen:2023aa} or through UV emission from AGN.

\subsection{Relation to other works}
\label{sec:Relation to other works}

In this work, we have argued for two distinct forms of starburst events at high redshift: internally driven bursts which are driven by the accretion of gas from the CGM and the ejection of that gas from the ISM in a breathing mode, and externally driven bursts which are driven by rapid and massive inflows from the IGM. We have also argued that the primary driver of galaxy burstiness at high redshift is the increased prevalence of externally driven bursts, caused by an increasing clumpy IGM and a higher rate of gas-rich mergers.

The breathing mode of star formation bursts has been identified in many previous works \citep{Christensen:2016aa,El-Badry:2016aa,Hopkins:2023aa,Semenov:2025aa}, and similarly galaxy interactions have long been identified as a cause of starbursts \citep[e.g.][]{Wilkinson:2018aa,Moreno:2021aa,Garay-Solis:2023aa}. The increased short-term variability of inflows from the IGM at high redshift has also been thought to play a role in increasing star-formation variability in the early Universe \citep{Tacchella:2018aa}. However, it has not been well understood how these different causes of burstiness relate to the extremely bursty behavior of galaxies observed recently by \textit{JWST} \citep[e.g.][]{Endsley:2024ab,Looser:2025aa}. We have identified large-scale rapid inflows as the dominant cause of increased burstiness in the Universe. However, we note that we have not distinguished between clumpy inflowing gas from the IGM and gas inflows due to mergers, as both can drive the externally driven mode by funneling gas rapidly into the ISM of a galaxy. 

\citet{Asada:2024aa} study a sample of 123 galaxies at $4.7<z<6.5$ and find that $\sim$60\% of their galaxies show evidence of bursty star formation and $\sim$40\% of their galaxies are interacting. They derive a strong connection between interacting galaxies and galaxies with bursty star formation, which may indicate that galaxy--galaxy interactions are an important source of the rapid gas inflows which drive the externally driven mode of starbursts. Additionally, \citet{Witten:2024aa} show that the surprising abundance of Lyman-$\alpha$ emitters at $z>7$ can be explained as starbursts induced by galaxy--galaxy interactions. However, while the merger rate has been shown to increase with redshift \citep{Duan:2025aa,Puskas:2025aa}, \citet{Fensch:2017aa} find that the high gas fraction mergers which are prevalent at high redshift may not be as efficient at driving starbursts. If this is the case, then the clumpiness of IGM gas inflows may be the dominant driver of externally driven bursts. We leave a detailed further investigation to future work.

Analytic frameworks to explain the burstiness of star formation at high redshift \citep[e.g.,][]{Faucher-Giguere:2018aa,Caplar:2019aa,Tacchella:2020aa,Furlanetto:2022aa} typically attribute burstiness in low-mass galaxies to the stochastic sampling of GMCs and in high-redshift galaxies to short dynamical times relative to the onset of supernovae feedback. Interestingly, \thzoom includes several channels of early stellar feedback \citep{Kannan:2025aa}, and \citet{Wang:2025aa} find little evidence of evolution in the GMC-scale SFE with redshift in \thzoom. This indicates that short dynamical times on the GMC scale do not necessarily increase cloud-scale SFE, and is further evidence that the increasing burstiness with redshift in \thzoom is indeed due to more variable gas supplies on the galaxy scale rather than cloud-scale physics. Our results suggest that while it is explicitly considered in some analytic modeling \citep[e.g.,][]{Caplar:2019aa,Tacchella:2020aa}, greater emphasis should be placed on the time variability of gas inflows.

It is also useful to compare our results to the study of starbursts in the local Universe. \citet{Cenci:2024aa} analyse galaxies in the FIREbox simulation \citep{Feldmann:2023aa} at low redshift (from $z=0$ to $z=1$) and make a similar distinction between modes of starburst. As their work is focused at low redshift, they do not consider clumpy IGM inflows and instead distinguish between interaction-driven starbursts and non-interaction-driven starbursts. They found that gas compaction occurs in non-interacting galaxies when the gas reservoir becomes gravitationally unstable, in agreement with previous work \citep{Dekel:2014aa,Danovich:2015aa,Zolotov:2015aa}. These non-interacting starbursts are similar to our internally driven mode, and although we have not directly explored the role of gas compaction in internally driven starbursts, in \citet{McClymont:2025ab} we show that the star formation in the \thzoom starburst galaxies is heavily centrally concentrated relative to the overall stellar population, which would align with gas-compaction driven internally driven bursts. This type of compaction is also seen in other simulations \citep{Tacchella:2016aa,Tacchella:2016ab,Lapiner:2023aa}. \citet{Cenci:2024aa} also found that galaxy interactions were important in driving starbursts at low redshift \citep[see also, for example,][]{Wilkinson:2018aa,Moreno:2021aa,Garay-Solis:2023aa}, and similarly to this work find that these starbursts are longer lasting and more extreme. However, they found that the role of galaxy interactions was still primarily through the compaction of gas, rather than the rapid supply of gas as seen in the externally driven mode of burstiness described in this work. This distinction highlights the increased importance of rapid gas supply through clumpy inflows and gas-rich mergers at high redshift compared to low redshift.

\section{Summary and Conclusions}
\label{sec:Conclusions}

In this work, we have investigated the star-forming main sequence (SFMS) at high redshift using the \thzoom\ simulations—a high-resolution extension of the \thesan\ project. These simulations incorporate detailed models of galaxy formation and evolution, including a multi-phase interstellar medium (ISM), local feedback around stellar particles, and on-the-fly radiative transfer using the {\sc arepo-rt} code. By post-processing the simulations with the \textsc{colt} radiative transfer code, we generated synthetic observables such as H$\alpha$ and UV luminosities, allowing for direct comparison with observational data.

Our primary goal was to analyze the SFMS and its scatter, focusing on how star formation rates (SFRs) and their variability depend on stellar mass and redshift. We fit the SFMS using both intrinsic SFRs averaged over different timescales and SFRs derived from H$\alpha$ and UV luminosities with a sample of galaxies across $3<z<11$. When studying SFMS scatter, we focus on galaxies with masses $10^7\,\mathrm{M}_{\odot} < \mathrm{M}_{\ast} < 10^{10}\,\mathrm{M}_{\odot}$. Our main conclusions are as follows:

\begin{itemize} 
\item \textbf{SFMS fits and normalization:} The intrinsic SFMS fits from our simulations align fairly well with theoretical expectations based on dark matter halo growth, with sSFR exhibiting a weak stellar mass dependence scaling as $\propto\mathrm{M_\ast}^{0.041\pm0.004}$ and a strong redshift dependence scaling as $\propto (1+z)^{2.64\pm0.03}$. At the high redshifts we consider here, the SFMS mass and redshift dependence are dependent on the SFR averaging timescale (quoted values are for SFR$_{10}$, see Tab.~\ref{tab:sfms_fits} for all fits).

\item \textbf{Impact of star-formation tracers:} When using SFRs derived from H$\alpha$ and UV luminosities, we observe systematic biases and scatter compared to intrinsic SFRs. H$\alpha$-derived SFRs tend to be underestimated due to factors like Lyman continuum (LyC) escape, while UV-derived SFRs are often overestimated and heavily scattered due to the assumption of constant star-formation history (SFH) over 100 Myr being at odds with the bursty nature of high-redshift galaxies. We tested which timescales are best probed by H$\alpha$ and UV emission, and provide revised SFR$_8$ and SFR$_{24}$ calibrations better suited for high-redshift galaxies with bursty star formation histories. The scatter in these calibrations can be added in quadrature to measurement errors to estimate uncertainty due to SFH variability, stellar metallicity, and complex physics such as LyC escape.

\item \textbf{Modes of starburst:} We identified two distinct modes of star formation burstiness contributing to the short-term scatter in the SFMS:
\begin{enumerate}
    \item \textit{Internally driven burstiness:} In this mode, galaxies experience short-lived bursts of star formation due to internal processes despite steady accretion of gas to the circumgalactic medium (CGM). Shortly after the onset of the burst, stellar feedback expels gas from the ISM and gas is rapidly depleted as it is converted into stars, which quenches the galaxy.
    \item \textit{Externally driven burstiness:} Here, variability in gas inflows at the CGM scale due to clumpy accretion or mergers leads to sustained periods of elevated star formation. These externally driven bursts can last significantly longer and are associated with massive inflows that fuel star formation while feedback is vented through lower-density regions of the CGM.
\end{enumerate}

\item \textbf{Outflows and gas dynamics:} Through cross-correlation analysis, we found strong correlations between SFR and gas inflow and outflow rates at the galactic scale. For the masses and redshifts we consider, this indicates that most bursts of star formation are preceded by rapid inflow of gas to the galaxy and followed by a rapid clearing of the ISM after a (median) $\sim$21\,Myr delay. There is a weaker correlation with gas inflow on the CGM scale, demonstrating that not all starbursts are associated with CGM-scale inflows.

\item \textbf{Scatter in the SFMS:} We analyzed the SFMS scatter as a function of SFR averaging timescale and found that the total scatter increases with decreasing redshift due to the increasing importance of longer-term environmental impacts on star formation.
%\item \textbf{Short-term scatter in the SFMS:} 
However, the SFMS scatter and CGM inflow scatter due to short-term variability (timescales less than 50\,Myr) increases with redshift. This suggests that galaxies are more bursty on short timescales in the early Universe, primarily due to the increasingly rapid variability of CGM-scale inflows at high redshift.

\item \textbf{Variability of star formation:} Our findings on the separate trends of short-term and total scatter with redshift imply that star formation is more variable on an individual galaxy level at high redshift, but more scattered on a population level at low redshift. Fig.~\ref{fig:ms_scheme} shows a schematic summarizing the arguments made in this work.

\item \textbf{Characterizing burst and quenching events:} By examining offsets from the SFMS, we quantified the durations and intensities of burst and quenching events. Both durations increase with stellar mass, with bursts (quenching events) lasting from $\sim$30\,Myr ($\sim$47\,Myr) in lower-mass galaxies to $\sim$38\,Myr ($\sim$83\,Myr) in higher-mass galaxies in our sample. Quenching events generally last longer than bursts, reflecting the self-regulating nature of starbursts compared to quenching events.

\item \textbf{UV scatter:} We find that intrinsic UV scatter converges to $\sigma_\mathrm{UV}\approx0.7\,\mathrm{mag}$ at $z>6$. This convergence is despite the increasingly bursty star formation at high redshift, and is due to the fact that longer-term scatter only appears with increasing cosmic time. While we find a decreasing trend of UV scatter with stellar mass at higher masses, for $M_\ast\lesssim 10^{8}\mathrm{M_\odot}$ the scatter is flat as a function of stellar mass. That we do not find increasing scatter with redshift or with decreasing stellar mass (for $M_\ast\lesssim 10^{8}\mathrm{M_\odot}$) may imply that the scatter alone cannot explain the over-abundance of bright galaxies seen at high redshift with \textit{JWST}, and that either increased scatter or increased UV luminosity (e.g. due to a top-heavy IMF or UV emission from AGN) is needed for low-mass galaxies at high redshift.

\end{itemize}

Our findings highlight the critical role of gas inflow variability at the CGM scale in driving the total scatter of the SFMS at lower redshifts and the increase in short-term burstiness of galaxies at high redshift. In the future, running simulations with detailed ISM and feedback modeling in larger volumes and down to $z=0$ will allow us to better characterize the transition from the extreme high-redshift regime of star formation to the ``slow boil'' regime seen in more massive galaxies at cosmic noon and in the local Universe.

\section*{Acknowledgements}

The authors are grateful to the referee for their helpful comments, which improved the manuscript. The authors gratefully acknowledge the Gauss Centre for Supercomputing e.V. (\url{www.gauss-centre.eu}) for funding this project by providing computing time on the GCS Supercomputer SuperMUC-NG at Leibniz Supercomputing Centre (\url{www.lrz.de}), under project pn29we. WM thanks the Science and Technology Facilities Council (STFC) Center for Doctoral Training (CDT) in Data Intensive Science at the University of Cambridge (STFC grant number 2742968) for a PhD studentship. WM and ST acknowledge support by the Royal Society Research Grant G125142. RK acknowledges support of the Natural Sciences and Engineering Research Council of Canada (NSERC) through a Discovery Grant and a Discovery Launch Supplement (funding reference numbers RGPIN-2024-06222 and DGECR-2024-00144) and York University's Global Research Excellence Initiative. EG is grateful to the Canon Foundation Europe and the Osaka University for their support through the Canon Fellowship. Support for OZ was provided by Harvard University through the Institute for Theory and Computation Fellowship. XS acknowledges the support from the National Aeronautics and Space Administration (NASA) grant JWST-AR-04814.

Various software packages were used in this work, including \textsc{numpy} \citep{Harris:2020aa}, \textsc{scipy} \citep{Virtanen:2020aa}, \textsc{matplotlib} \citep{Hunter:2007aa}, and \textsc{astropy} \citep{Astropy-Collaboration:2013aa,Astropy-Collaboration:2018aa,Astropy-Collaboration:2022aa}.

%%%%%%%%%%%%%%%%%%%%%%%%%%%%%%%%%%%%%%%%%%%%%%%%%%
\section*{Data Availability}

All simulation data, including snapshots, group, and subhalo catalogs and merger trees will be made publicly available in the near future. Data will be distributed via \url{www.thesan-project.com}. Before the public data release, data underlying this article will be shared on reasonable request to the corresponding author(s).

%%%%%%%%%%%%%%%%%%%% REFERENCES %%%%%%%%%%%%%%%%%%

% The best way to enter references is to use BibTeX:

\bibliographystyle{mnras}
\bibliography{main} % if your bibtex file is called example.bib

% Alternatively you could enter them by hand, like this:
% This method is tedious and prone to error if you have lots of references
%\begin{thebibliography}{99}
%\bibitem[\protect\citeauthoryear{Author}{2012}]{Author2012}
%Author A.~N., 2013, Journal of Improbable Astronomy, 1, 1
%\bibitem[\protect\citeauthoryear{Others}{2013}]{Others2013}
%Others S., 2012, Journal of Interesting Stuff, 17, 198
%\end{thebibliography}

%%%%%%%%%%%%%%%%%%%%%%%%%%%%%%%%%%%%%%%%%%%%%%%%%%

%%%%%%%%%%%%%%%%% APPENDICES %%%%%%%%%%%%%%%%%%%%%

\appendix

\section{Impact of centrals and satellites}
\label{sec:Impact of centrals and satellites}

\begin{figure} 
\centering
	\includegraphics[width=\columnwidth]{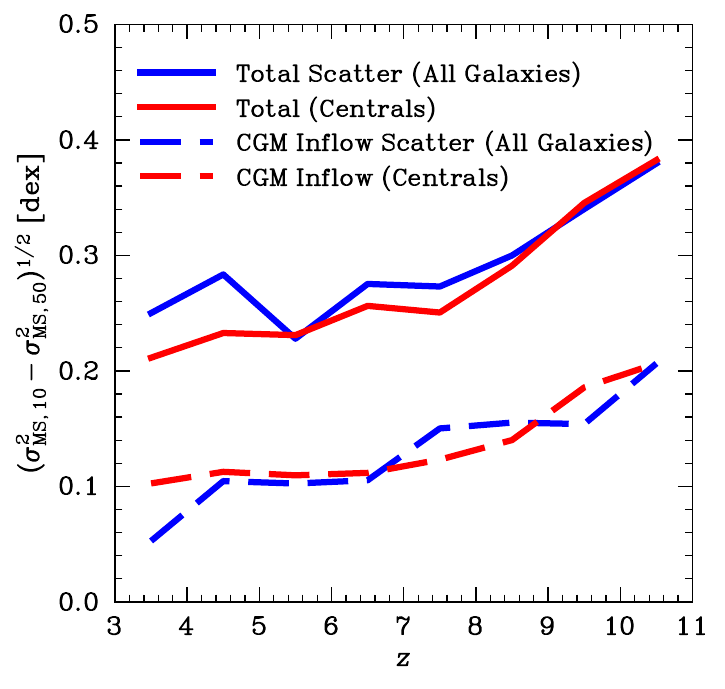}
    \caption{The same as Fig.~\ref{fig:feld_ratios} except only including central galaxies in the analysis (red lines). For comparison, the blue lines are the same as in Fig.~\ref{fig:feld_ratios} (i.e. are the result of jointly fitting centrals and satellites).}
    \label{fig:feld_ratios_cent}
\end{figure}

\begin{figure} 
\centering
	\includegraphics[width=\columnwidth]{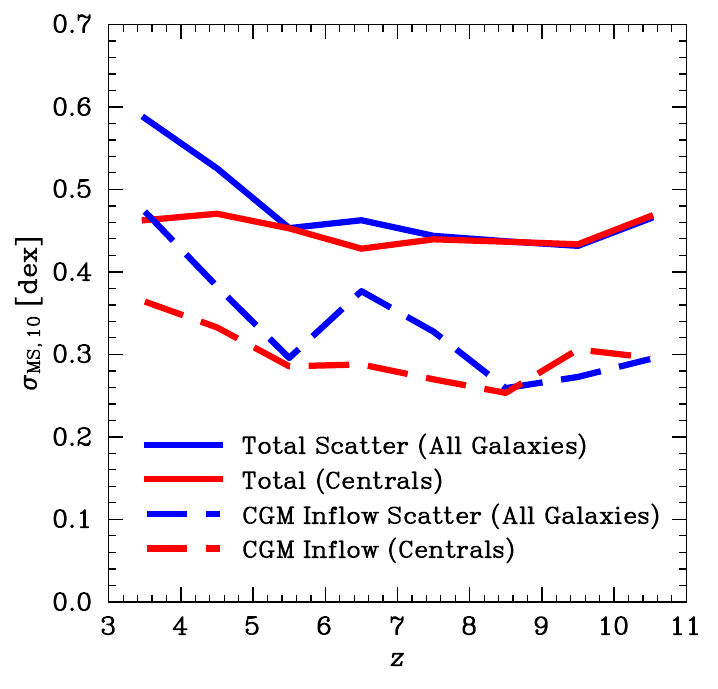}
    \caption{The same as Fig.~\ref{fig:vir_inflow_ratios} except only including central galaxies in the analysis (red lines). For comparison, the blue lines are the same as in Fig.~\ref{fig:vir_inflow_ratios} (i.e. are the result of jointly fitting centrals and satellites). The redshift evolution of the scatter is much shallower when only centrals are included. This is because centrals are less sensitive to environmental effects, as they are, by definition, the most massive galaxy in their local area.}
    \label{fig:vir_inflow_ratios_cent}
\end{figure}

To understand the impact that satellite galaxies have on the SFMS and its scatter, we also repeated the analysis only including central galaxies. Central galaxies are defined as the most massive subhalo in a FoF halo. All other selection criteria remain the same, and our centrals-only sample includes 14685 subhaloes, comprised of 313 unique trees. In Tab.~\ref{tab:sfms_fits_cent}, we show a selection of SFMS fits to the sample of central galaxies. 

In Fig.~\ref{fig:feld_ratios_cent} and Fig.~\ref{fig:vir_inflow_ratios_cent} we show the scatter calculated for centrals only as the red solid line. For clarity, we have re-run the entire analysis with a centrals-only selection to calculate the scatter, including using the centrals-only SFMS fits shown in Tab.~\ref{tab:sfms_fits_cent}. The blue lines are the same as in Fig.~\ref{fig:feld_ratios} and Fig.~\ref{fig:vir_inflow_ratios} (i.e. are the result of jointly fitting centrals and satellites). This allows for an easier direct comparison of the changes when only centrals are included. We note that the double power law fits do not well capture the form of the redshift evolution for the centrals-only scatter, so we do not include them in the figures. The scatter for centrals only evolves significantly less with redshift for the centrals-only sample, as seen in Fig.~\ref{fig:vir_inflow_ratios}. This is because centrals are less subject to environmental effects than satellites due to the fact that they are the most massive galaxy in their neighborhood. In Fig.~\ref{fig:feld_ratios} we show that the increase in the short-term scatter still holds for the centrals-only sample, as expected. In fact, the increase is more obvious, likely because the overall scatter is lower and therefore is easier to disentangle in quadrature. The short-term CGM inflow scatter still well tracks the short-term SFMS scatter.

We also used this as an opportunity to test how well subhalo particles were selected for our outflow analysis. Subhalo membership, especially out to large radii such as those that we consider for the CGM scale. Therefore, we repeated our analysis with a new approach for selecting gas particles. Rather than only including subhalo particles, we select particles from the FoF group. This selection is purely spatial and therefore less vulnerable to subhalo membership issues, especially on the halo scale. When we carried out the fits with these new outflow catalogues, we reproduced the trends from Fig.~\ref{fig:feld_ratios_cent} and Fig.~\ref{fig:vir_inflow_ratios_cent} with no significant differences, increasing our confidence in our full centrals and satellites analysis.

\begin{table}
    \centering
    \begin{tabular}{ccccc}
        \hline
        \multicolumn{5}{|c|}{Redshift and mass-dependent SFMS fit (centrals only)} \\
        \hline
        Tracer & $\mathrm{s_b}$ & $\beta$ & $\mu$ & \\
        SFR$_{10}$ & $0.073\pm0.005$ & $0.029\pm0.005$ & $2.24\pm0.03$ & \\
        SFR$_{50}$ & $0.090\pm0.005$ & $0.029\pm0.003$ & $2.09\pm0.02$ & \\
        SFR$_{100}$ & $0.131\pm0.006$ & $0.021\pm0.003$ & $1.84\pm0.03$ & \\
        \hline
    \end{tabular}
    \caption{The same as Tab.~\ref{tab:sfms_fits} except we provide for fits only including central galaxies in our fits. }
    \label{tab:sfms_fits_cent}
\end{table}

\section{Different measures of scatter}
\label{sec:Different measures of scatter}

\begin{figure*} 
\centering
	\includegraphics[width=\textwidth]{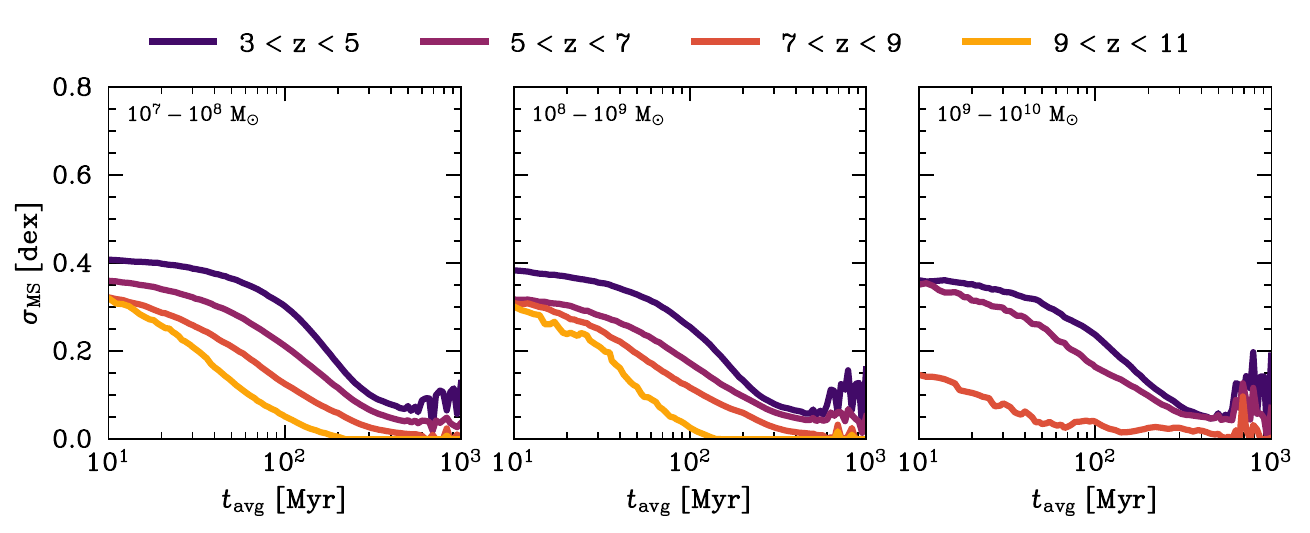}
    \caption{The same as Fig.~\ref{fig:sfms_scat} except we have measured the scatter by only considering galaxies above the SFMS ($\Delta\mathrm{MS}>0$). To measure the scatter we mirror the sSFRs around $\Delta\mathrm{MS}=0$ and fit a normal distribution. The scatter measured this way is less reliable at longer timescales (see text for details).}
    \label{fig:sfms_scat_above}
\end{figure*}

\begin{figure*} 
\centering
	\includegraphics[width=\textwidth]{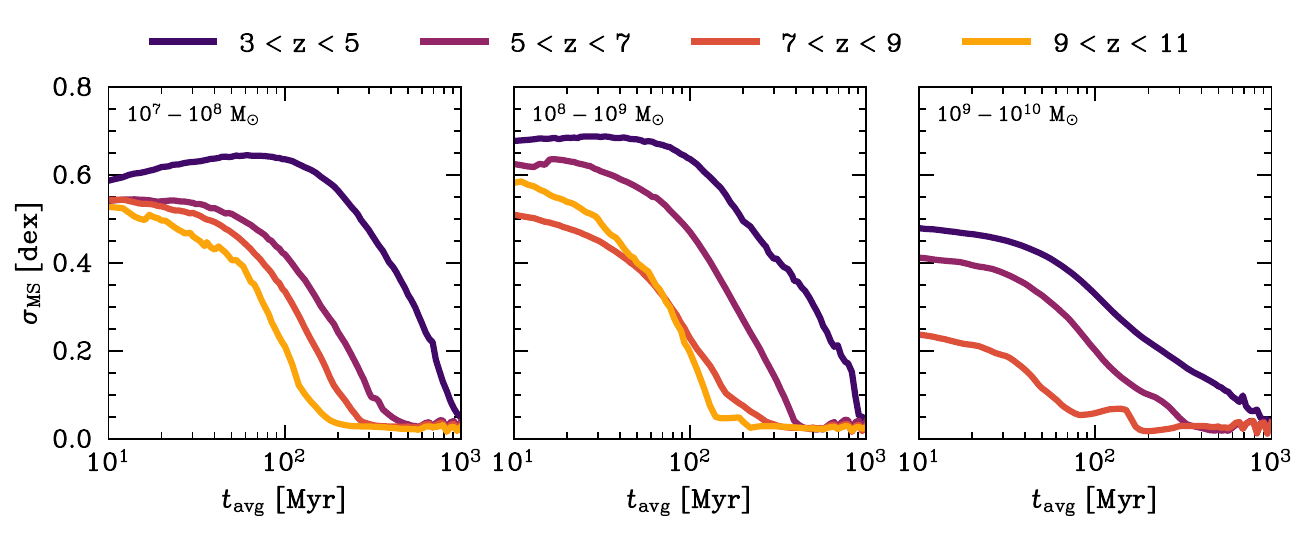}
    \caption{The same as Fig.~\ref{fig:sfms_scat} except we have measured the scatter simply by fitting a normal distribution. Scatter measured in this way does not accurately measure the scatter or capture trends with increasing $t_\mathrm{avg}$ because the SFRs are not log-normally distributed.}
    \label{fig:sfms_scat_norm}
\end{figure*}

It is hard to accurately measure the SFRs of galaxies in the tail of the SFR distribution, even if a survey is deep enough to detect them. For this reason, in Fig.~\ref{fig:sfms_scat_above} we provide fits to the SFMS scatter where we exclusively consider galaxies with $\Delta\mathrm{MS}>0$. We measure the scatter by mirroring the sSFRs around $\Delta\mathrm{MS}=0$ and then fit a normal distribution. This scatter is easier to assess observationally because it only requires an accurate fit to the peak of the SFR distribution and a complete sample of galaxies above the SFMS, rather than requiring SFRs to be measured accurately for a mass-complete sample of galaxies.

We caution that the scatter measured with this approach is less reliable, especially for longer timescales ($t_\mathrm{avg}\gtrsim100$\,Myr), because it is sensitive to the accurate fitting of the SFMS itself. For our fits, the SFMS fit on longer timescales is less accurate because we get to the regime where, for high-redshift galaxies, all of the stellar mass has been formed within $t_\mathrm{avg}$. This causes a breakdown of the redshift scaling we have assumed. This does not have a strong impact on our fiducial fits because they are independent of normalization.

In Fig.~\ref{fig:sfms_scat_norm} we show the scatter as measured simply by fitting a normal distribution to $\Delta\mathrm{MS}$. These should not be used for any purpose, but demonstrate the necessity of recognizing the non-log-normal distribution of SFRs when fitting the SFMS scatter.

%%%%%%%%%%%%%%%%%%%%%%%%%%%%%%%%%%%%%%%%%%%%%%%%%%

% Don't change these lines
\bsp	% typesetting comment
\label{lastpage}
\end{document}